\documentclass[10pt,twocolumn,longbibliography,amsmath,amssymb,floatfix,superscriptaddress,aps,prl]{revtex4-2}

\usepackage[english]{babel}
\usepackage{graphicx}
\usepackage{subfigure}
\usepackage{subcaption}
\usepackage{amsmath}  
\usepackage{overpic}
\usepackage{caption}
\usepackage{ragged2e}
\usepackage{caption}
\usepackage{tikz}
\usepackage{tikz-3dplot}
\usepackage{graphicx}

\usepackage{dcolumn}
\usepackage{bm}

\usepackage{xcolor}  
\usepackage{hyperref} 

\definecolor{myblue}{rgb}{0.0, 0.0, 1.0} 

\hypersetup{
    colorlinks=true,   
    linkcolor=myblue,  
    citecolor=myblue,  
    urlcolor=myblue    
}

\begin{document} 

\preprint{APS/123-QED}

\title{All-Optical Generation of Dense, Multi-GeV, Longitudinally-Polarized Positron Beams}

\author{Rui-Qi Qin $^{\dag}$}
\affiliation{Department of Nuclear Science and Technology, Xi'an Jiaotong University, Xi'an 710049, China}\thanks{These authors contributed equally to this work}%
\affiliation{Northwest Institute of Nuclear Technology, Xi’an 710024, China}

\author{Peng-Pei Xie $^{\dag}$}
\affiliation{Department of Nuclear Science and Technology, Xi'an Jiaotong University, Xi'an 710049, China}

\author{Yan-Fei Li}\email{liyanfei@xjtu.edu.cn}    
\affiliation{Department of Nuclear Science and Technology, Xi'an Jiaotong University, Xi'an 710049, China}

\author{Xian-Zhang Wu}
\affiliation{Department of Nuclear Science and Technology, Xi'an Jiaotong University, Xi'an 710049, China}

\author{Zheng-Yang Zuo}
\affiliation{Department of Nuclear Science and Technology, Xi'an Jiaotong University, Xi'an 710049, China}

\author{Bing-Jun Li}
\affiliation{Department of Nuclear Science and Technology, Xi'an Jiaotong University, Xi'an 710049, China}
\affiliation{Northwest Institute of Nuclear Technology, Xi’an 710024, China}

\author{Jun Liu}\email{liujun@nint.ac.cn}
\affiliation{Northwest Institute of Nuclear Technology, Xi’an 710024, China}

\author{Liang-Liang Ji}
\affiliation{State Key Laboratory of High Field Laser Physics,
Shanghai Institute of Optics and Fine Mechanics,
Chinese Academy of Sciences, 
Shanghai 201800, China}

\author{Yu-Tong Li}  
\affiliation{Beijing National Laboratory for 
Condensed Matter Physics, Institute of Physics, CAS, Beijing 100190,
China}
\affiliation{School of Physical Sciences, University of Chinese Academy of Sciences, Beijing 100049, China}
\affiliation{Collaborative Innovation Center of IFSA (CICIFSA), Shanghai Jiao Tong University, Shanghai 200240, China}
\affiliation{Songshan Lake Materials Laboratory, Dongguan, Guangdong 523808, China}


\date{\today}

\begin{abstract}

The production of high-yield, longitudinally polarized positron beams represents an outstanding challenge in advanced accelerator science. Laser-driven schemes offer a compact alternative but typically yield only transverse polarization, or require pre-polarized electron beams, and struggle to efficiently accelerate positrons to high energies. Here, we introduce an all-optical scheme that overcomes these limitations by integrating positron generation, acceleration, and spin manipulation in a unified framework. Through a head-on collision between an ultraintense, circularly polarized laser pulse and a counterpropagating unpolarized electron beam, we drive a robust QED cascade. The nonlinear Breit-Wheeler process within the cascade produces positrons that are born directly within the strong laser field. Crucially, these positrons are instantaneously captured and accelerated to multi-GeV energies (up to $\sim$9 GeV) via a direct laser acceleration mechanism, while their spins are simultaneously rotated to longitudinal alignment by the field dynamics. Our Monte-Carlo simulations confirm the simultaneous achievement of a high positron yield ($\sim$20 $e^+/e^-$), a high average longitudinal polarization ($\sim$50\%), and GeV-scale energies. This all-optical source, feasible at upcoming ultraintense laser facilities, presents a compact and efficient solution for applications in collider physics and fundamental high-energy experiments.

\end{abstract}

\maketitle


Longitudinally spin-polarized positrons, with spin aligned parallel to momentum, are indispensable for probing fundamental symmetries and testing the Standard Model in high-energy physics. Their helicity sensitivity enables precision studies of electroweak couplings at future colliders such as the International Linear Collider (ILC) \cite{bambade2019international, Kovalenko2012SpinTracking}, investigations of chiral dynamics in QED cascades \cite{Seipt2021PolarizedQEDCascades}, and searches for CP-violating physics beyond the Standard Model \cite{moortgat2008polarized}. These endeavors demand high-flux positron sources delivering GeV beams with high longitudinal polarization ($>30\%$) and substantial yields ($> 1.5$ $e^+/e^-$) \cite{richard2001tesla, moortgat2008polarized}.

Conventional positron polarization techniques, however, are constrained by a fundamental trade-off among yield, polarization degree and efficiency. Radioactive $\beta^+$ sources produce beams of low intensity and high divergence \cite{zitzewitz1979spin}.Synchrotron-based methods, such as the Sokolov-Ternov effect, require hour-long polarization buildup times \cite{sokolov1964polarization}. The most advanced approach, helicity transfer from circularly polarized $\gamma$ rays via Bethe-Heitler pair production in solid targets, is fundamentally limited by depolarization from multiple Coulomb scattering \cite{potylitsin1997production}. This results in theoretical yields below $0.01$ $e^+/e^-$ \cite{olsen1959photon} and experimental yields as low as $\sim 10^{-5}$ $e^+/e^-$ \cite{Abbott2016, Omori2006}, primarily due to limitations in the photon spectrum and yield-polarization trade-offs during collimation.

Laser-driven all-optical schemes have emerged as promising alternatives, leveraging ultraintense fields to generate positrons via the nonlinear Breit-Wheeler (NBW) process in asymmetric laser setups \cite{chen2019polarized, wan2020ultrarelativistic, zhuang2023laser}. While these methods enhance yield and efficiency, they typically produce only transversely polarized positrons with limited charge ($\lesssim 1$ $e^+/e^-$). Proposals employing solid targets with 100-PW-class lasers \cite{xue2023generation,song2022dense} or ultra-dense electron beams \cite{zhu2024dense} could achieve nanocoloumb-level yields but still fail to produce the essential longitudinal polarization required for applications \cite{richard2001tesla,moortgat2008polarized,Flottmann1993,duan2019}. Polarization rotators \cite{Steffens1993,Buon1986, Li2025} offer a potential solution but introduce significant complexities, including large footprints and substantial particle losses from energy-bandwidth mismatch.

An alternative helicity-transfer scheme \cite{li2020production} could directly yield longitudinal polarization but requires pre-polarized GeV electron beams, a capability yet to be demonstrated experimentally in all-optical setups \cite{wen2019,li2019ultrarelativistic}. Furthermore, accelerating positrons to GeV energies in plasma wakefields remains challenging due to their inherent defocusing in electron-driven wakes \cite{Liu2022}, unlike electrons which are efficiently accelerated \cite{Lu2006,RevModPhys2009}. Consequently, the generation of dense, multi-GeV, longitudinally polarized positron beams from unpolarized electrons remains an outstanding challenge for all-optical approaches.

\begin{figure}[t] 
    \includegraphics[width=1.0\linewidth]{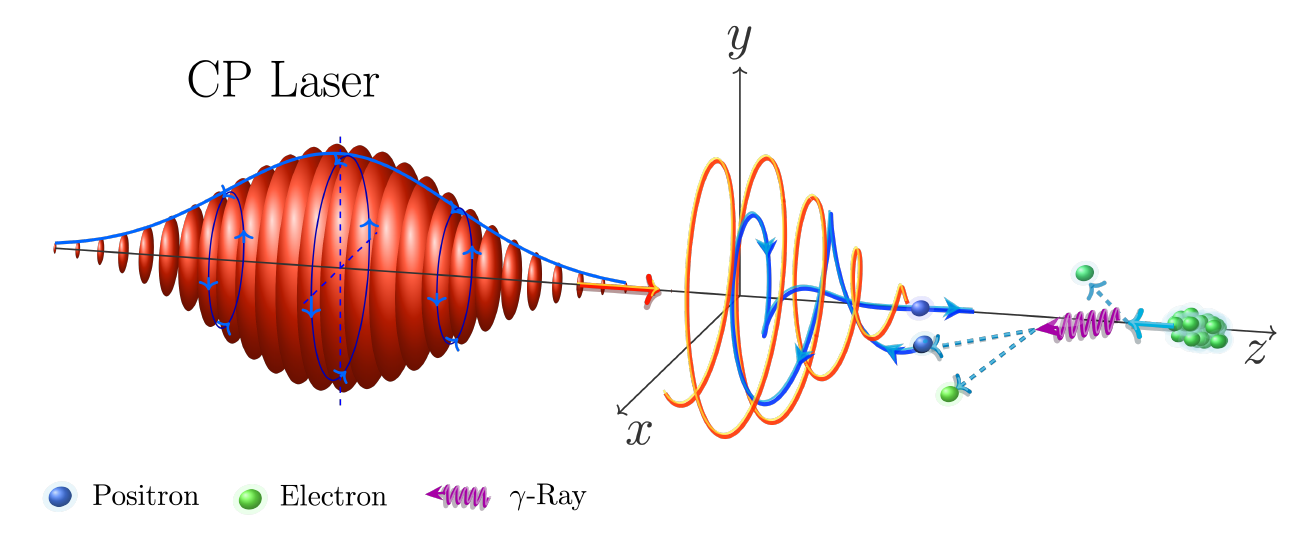}
\caption{\justifying Generation of longitudinally polarized positrons via collision of a counterpropagating ultraintense laser pulse and unpolarized electron beam. NBW pair production generates $e^+e^-$ pairs from $\gamma$-photons created by NCS. Newborn positrons, initially propagating along $-z$, are reflected and backward-accelerated by the laser field. During their spiral trajectory, spin precession induced by the phase-dismatched field efficiently converts transverse polarization into longitudinal polarization.}
\label{fig1}
\vskip -0.3 cc
\end{figure}

In this Letter, we present a compact, all-optical scheme that simultaneously generates, accelerates, and polarizes positrons in a single stage. We demonstrate that a head-on collision between a circularly polarized, ultraintense laser pulse and a counterpropagating, unpolarized electron beam directly yields longitudinally polarized positrons with high yield ($\sim 20$ $e^+/e^-$), GeV-scale energies (up to $\sim 9$ GeV), and substantial polarization ($\sim 50\%$ on average, peaking above $70\%$). The mechanism initiates with nonlinear Compton scattering (NCS), producing abundant high-energy photons, followed by NBW pair creation, which imparts an initial transverse polarization to the $e^+e^-$ pairs. Crucially, the positrons are born within the strong-field vacuum of the laser pulse. Radiation reaction rapidly damps their transverse momentum, facilitating reflection and trapping by the laser field. Subsequent direct laser acceleration (DLA) \cite{Pukhov1999,Gonoskov2022,Martinez2023} via the ${\bm v}\times{\bf B}$ force sustains multi-GeV energy gain. Simultaneously, the phase-mismatched field conditions induce rapid spin precession, quantitatively converting the initial transverse polarization into a longitudinal alignment. This intrinsic integration of all key functionalities—without requiring pre-polarized beams, external converters, or complex beamlines—establishes a practical pathway toward compact polarized positron sources for next-generation high-energy physics.

A Monte Carlo method \cite{li2020production,Wu2025} that self-consistently incorporates polarization effects in strong-field QED is employed to simulate the polarization dynamics of photons, electrons, and positrons throughout the interaction. This approach leverages the local constant field approximation (LCFA) \cite{li2019ultrarelativistic,li2020polarized,baier1998electromagnetic,ritus1985quantum,di2018implementing,di2019improved}, whose validity is well-established for ultraintense laser fields characterized by $a_0 \equiv |e|E_0/(m\omega_0) \gg 1$ \cite{baier1998electromagnetic,ritus1985quantum}. Within the LCFA framework, the probability of photon emission (and subsequent pair production) is governed by the quantum strong-field parameter $\chi_{e,\gamma} \equiv |e|\sqrt{-(F_{\mu\nu}p^{\nu})^{2}}/m^{3}$ \cite{ritus1985quantum}. Here, $E_0$ and $\omega_0$ denote the amplitude and frequency of the laser field, respectively; $e$ and $m$ are the electron charge and mass; $F_{\mu\nu}$ is the electromagnetic field tensor; and $p^{\nu}$ is the four-momentum of the electron (or photon for pair production). Relativistic units ($c=\hbar=1$) are used throughout this work.

\begin{figure}[t]
    \centering
    \begin{minipage}[b]{\linewidth}
        \centering
        \includegraphics[width=1.0\linewidth]{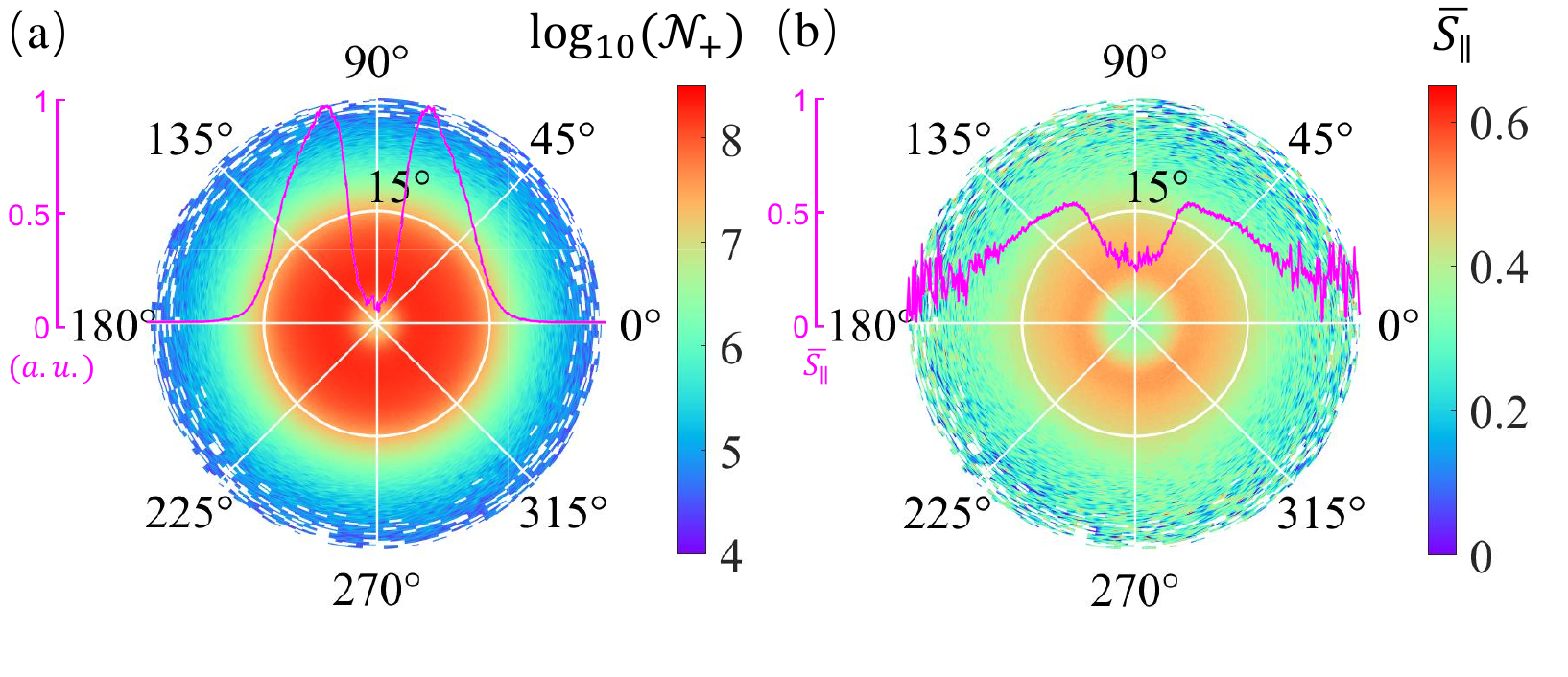}
 \label{fig:2(a-b)}
    \end{minipage}
    \vskip -1.6 cc    
    \begin{minipage}[b]{\linewidth}
        \centering
        \includegraphics[width=1.0\linewidth]{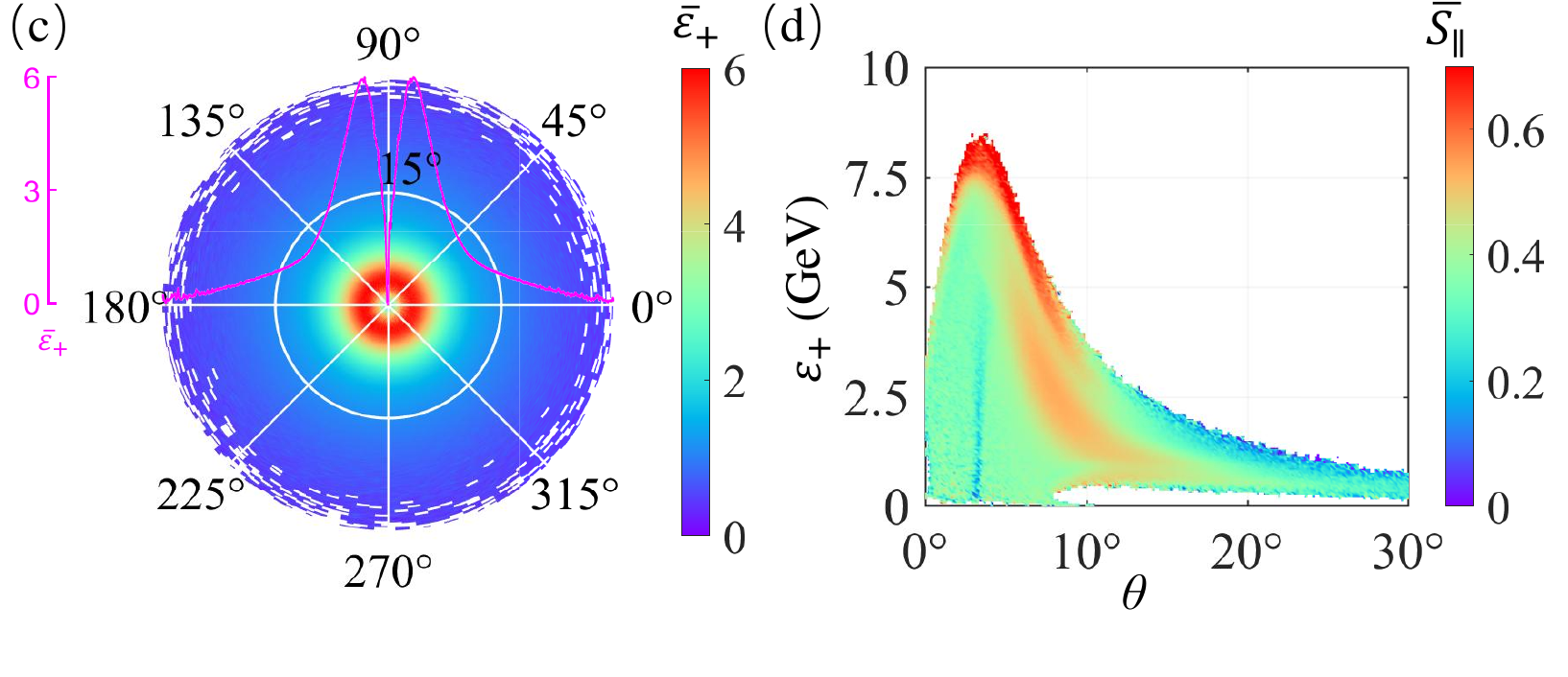}
         \label{fig:2(c-d)}
    \end{minipage}
     \vskip -1.7 cc
         \begin{minipage}[b]{\linewidth}
        \centering
        \includegraphics[width=1.0\linewidth]{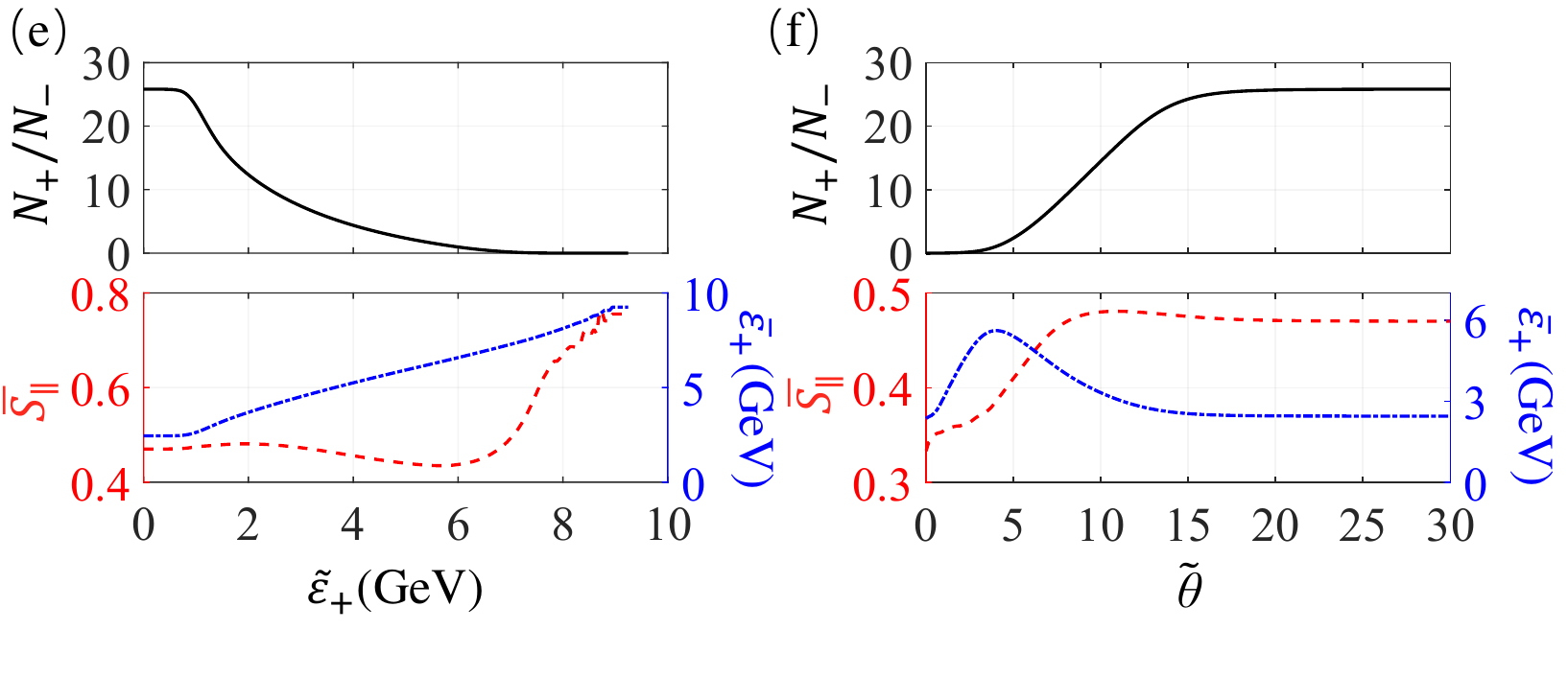}
         \label{fig:2(e)}
    \end{minipage}
     \vskip -1.45 cc
     \caption{\justifying (a)Angular distribution of number density $\log_{10}(\mathcal{N}_{+})$, with $\mathcal{N}_{+}=dN_{e^+}/({\rm sin}{\theta}d\theta d\phi)$ , (b) average longitudinal polarization $\bar{S}_{\parallel}$ and (c) average energy $\bar\varepsilon_+$ of positrons vs $\theta$ $\in [0,30^\circ]$ and $\phi$ $\in [0,360^\circ]$. Here, $\theta$ and $\phi$ are the polar and azimuthal angles with respect to $+z$ axis. Purple curves are normalized density, $\bar{S}_{\parallel}$ and $\bar\varepsilon_{+}$ with respect to $\theta$ along $\phi=0, 180^\circ$. (d)Distribution of $\bar{S}_{\parallel}$ vs $\varepsilon_{+}$ and $\theta$. (e) Positron yield (black solid), average polarization (red dashed), and average energy (blue dash-dotted) as functions of the lower energy cutoff $\tilde{\varepsilon}_+$; (f) the same quantities as functions of the angular acceptance $\tilde{\theta}$. The energy selection corresponds to collecting positrons with energies from $\tilde{\varepsilon}_+$ to 10 GeV, while the angular selection corresponds to collecting positrons within a polar angle from $0^\circ$ to $\tilde{\theta}$.
     }
    \label{fig:2}      
 \vskip -0.3 cc
\end{figure}

A representative result for polarized positron generation is presented in Fig.~\ref{fig:2}. The simulation involves a counter-propagation setup between an ultra-intense, tightly focused, circularly polarized Gaussian laser pulse and an ultra-relativistic electron beam. The laser pulse is characterized by the following parameters: a peak intensity of $I_0 \approx 2.76 \times 10^{24}~\text{W/cm}^2$ (corresponding to $a_0 = 1000\sqrt{2}$), a wavelength $\lambda_0 = 1~\mu\text{m}$, a focal radius $w_0 = 5\lambda_0$, and a pulse duration $\tau = 5T_0$, where $T_0$ is the optical period. The incident electron beam is configured with parameters typical of a laser-wakefield-accelerated source \cite{RevModPhys2009, gonsalves2019petawatt, leemans2014multi}. It is initially unpolarized, with a mean spin polarization vector $(\bar{S}_x, \bar{S}_y, \bar{S}z) = (0, 0, 0)$, a mean energy $\varepsilon_0 = 6$ GeV, a relative energy spread $\Delta{\varepsilon_0}/\varepsilon_0 = 6\%$, and an angular divergence $\Delta\theta = 0.2$ mrad. The bunch contains $N_e = 1 \times 10^6$ electrons, distributed uniformly over a longitudinal length $L_e = 6\lambda_0$, and possesses a transverse Gaussian profile with waist $w_e = 2\lambda_0$ and standard deviations $\sigma_x = \sigma_y = 0.6\lambda_0$. We emphasize that substantial polarized positron yields are also achievable at lower, more accessible laser intensities, as shown in Fig.~\ref{fig:5}(a) and  the Supplemental Material \cite{SM}. The experimental feasibility is supported by current laser technology capable of reaching intensities above $10^{23}~\text{W/cm}^2$ \cite{yoon2021realization}, with 100-PW-class facilities underway \cite{danson2019petawatt}. Moreover, laser wakefield acceleration can now deliver multi-GeV electron beams \cite{leemans2014multi, gonsalves2019petawatt}. 



\begin{figure}[h]
    \centering
     \vskip -0.3 cc
    \begin{minipage}[b]{\linewidth}
        \centering
        \includegraphics[width=1.0\linewidth]{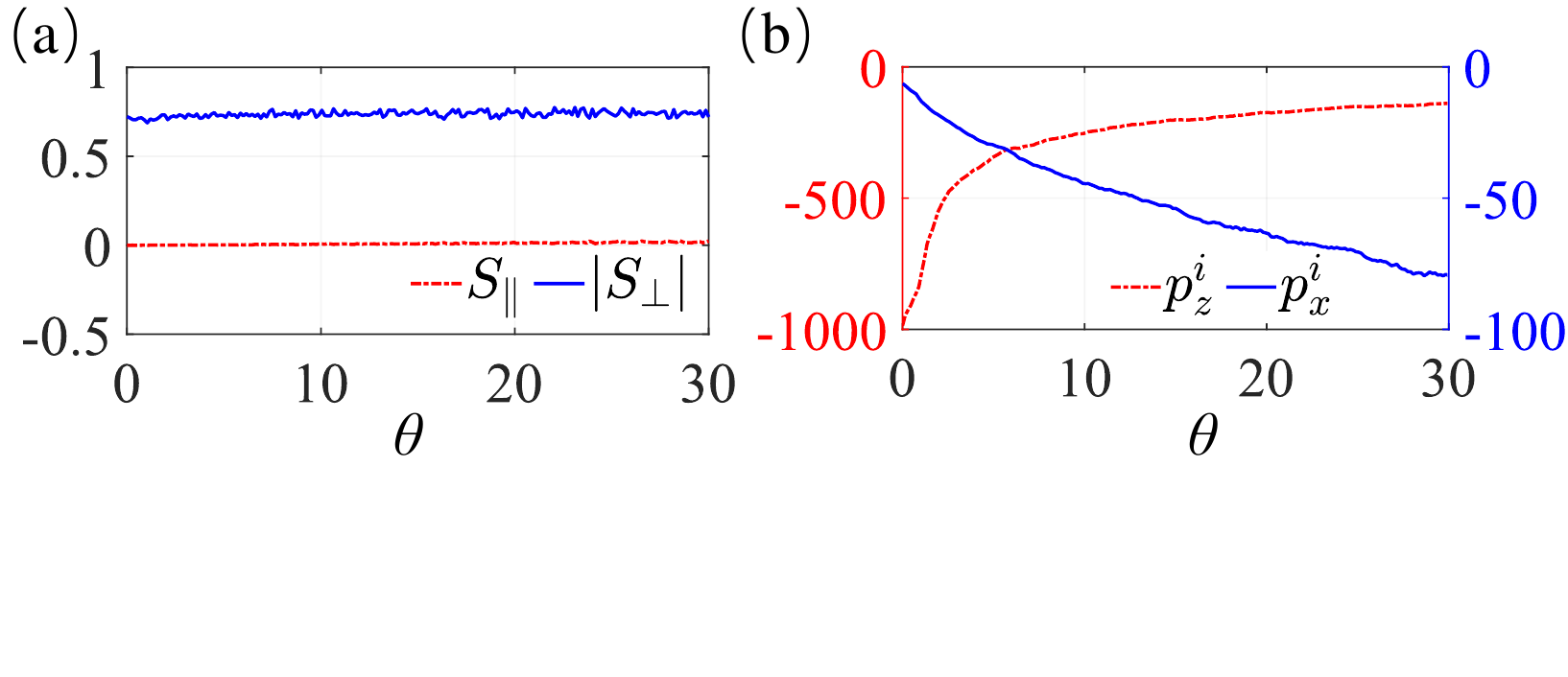}
         \label{}
    \end{minipage}
      \vskip -3.5 cc
     \begin{minipage}[b]{\linewidth}
        \centering
        \includegraphics[width=1.0\linewidth]{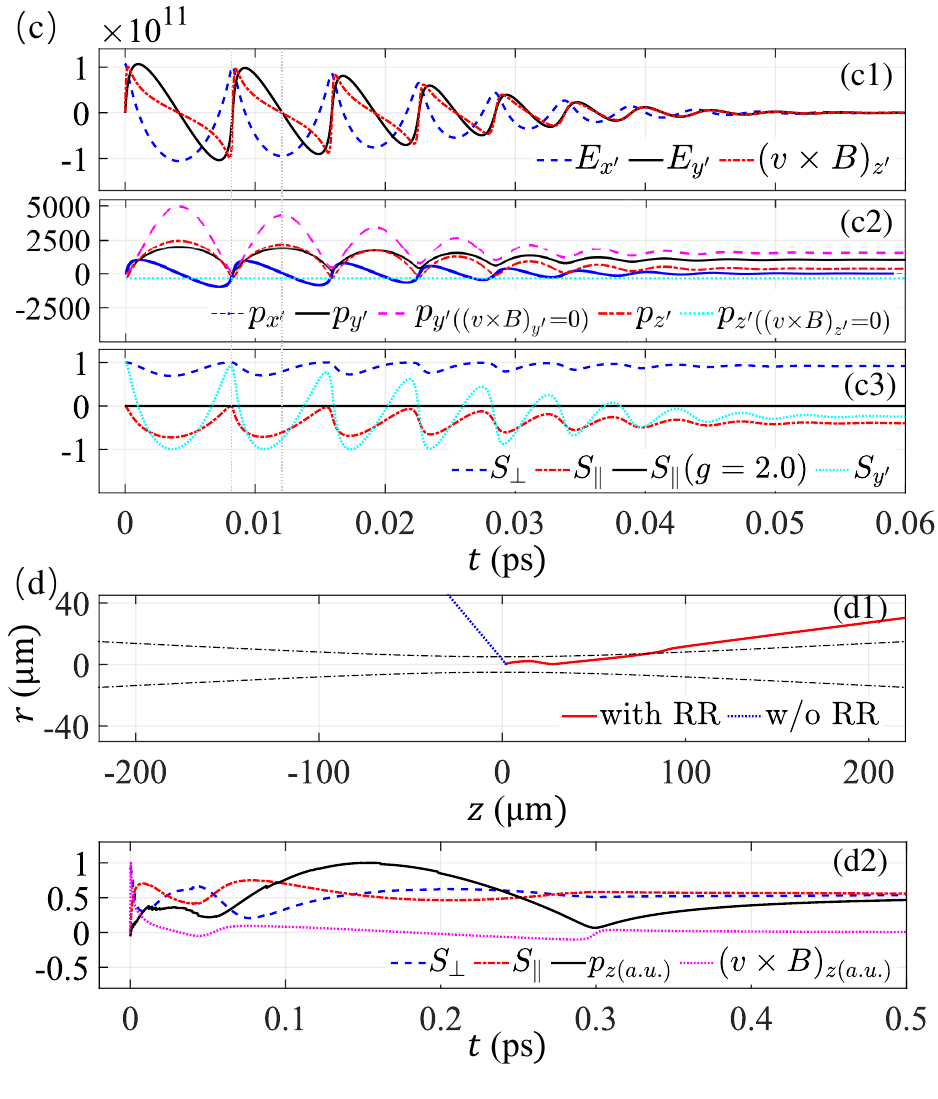}
         \label{}
    \end{minipage}

     \vskip -1.6 cc
    \caption{\justifying
   (a) Initial longitudinal \({S}_{\parallel}^i\) and transverse \({|{S}_{\perp}^i|}\) polarization of a positron at birth versus $\theta$ at $\phi=0^\circ$. (b) Corresponding initial momentum components $\bar{{p}_{z}^i}$, $\bar{{p}_{x}^i}$ (in units of $mc$). (c) Temporal evolution for a representative positron (radiation reaction neglected) in a plane-wave laser field: (c1) laser field components \(E_{x^{'}}\), \(E_{y^{'}}\), and the longitudinal Lorentz force component of \(({\bm v}\times {\bf B})_{z^{'}}\);  (c2) momentum components \(p_{x^{'}}\), \(p_{y^{'}}\), \(p_{z^{'}}\); (c3) polarization components \(S_\perp\), \({S}_{\parallel}\) (for physical and $g=2$ \(g\)-factor), and \(S_{y^{'}}\). (d) Spatial and dynamic evolution in the realistic laser field:  (d1) spatial trace $r=\sqrt{(x^2+y^2)}$ with the laser focus profile (black dash-dotted lines); (d2) dynamics of  ${|{S}_{\perp}|}$, ${S}_{\parallel}$, $p_z$ and \(({\bm v}\times {\bf B})_{z}\).
      } \label{fig3}
    
     \vskip -0.6 cc
\end{figure}

Figures\ref{fig:2}(a)–(c) present the angular distributions of the reflected positron number density, along with their average longitudinal polarization $\bar{S}_{\parallel}$ and mean energy $\bar{\varepsilon}_{+}$. Positrons are predominantly distributed within the polar angle range $\theta \in [0,30^\circ]$, with a total yield of $2.58 \times 10^7$ (equivalent to 25.8 $e^+$/$e^-$), an average polarization of 46.8\%, and a mean energy of 2.45 GeV.  A pronounced annular region between $\theta = 5^\circ$ and $12^\circ$ exhibits enhanced values of both $\bar{S}_{\parallel}$ and $\bar{\varepsilon}_{+}$ [purple curves in Figs. \ref{fig:2}(b) and (c)], reaching 48.9\% and 2.56 GeV on average, respectively. Given the relatively low particle density at smaller angles, including the central region has only a minor effect on overall polarization. Therefore, for practical experimental applications, it is beneficial to collect positrons within a broader angular range. For instance, selecting positrons within $\theta \in [0, 12^\circ]$ results in a beam with 48.0\% polarization, a total yield of 19.4 e$^+$/e$^-$, and a mean energy of 2.87 GeV. 

For clarity, Fig. \ref{fig:2}(d) illustrates the average longitudinal polarization $\bar{S}_\parallel$ as a function of both energy $\varepsilon_+$ and angle $\theta$. Positron energies extend up to 9 GeV, with $\bar{S}_{\parallel}$ increasing markedly at higher energies—reaching up to 72.7\%, at  $\theta \approx 3^\circ$. The dependence of $\bar{S}_\parallel$ on both $\theta$ and $\varepsilon_{+}$ enables the extraction of highly longitudinally polarized positron beams through post-selection in energy and angle, as shown in Figs. \ref{fig:2}(e) and (f).  For example, selecting positrons with energies above 7.6 GeV yields $\bar{S}_\parallel = 60.2\%$ and a normalized yield of $0.022$ $e^+$/$e^-$, while those above 7.0 GeV achieve $\bar{S}_\parallel = 51.2\%$ with a yield of $0.118$ $e^+$/$e^-$. Similarly, selecting positrons within a collection angle of $\tilde{\theta} = 10^\circ$ yields $\bar{S}_\parallel = 48.0\%$, a normalized yield of $14.5$ $e^+$/$e^-$ and a mean energy of 3.30 GeV, whereas a broader angle of $\tilde{\theta} = 15^\circ$ results in $\bar{S}_\parallel = 47.5\%$ with a yield of $24.3$ $e^+$/$e^-$ and a mean energy of 2.54 GeV. 

\begin{figure}[htb]
    \centering
     \vskip -0.3 cc
    \begin{minipage}[b]{\linewidth}
        \centering
        \includegraphics[width=1.0\linewidth]{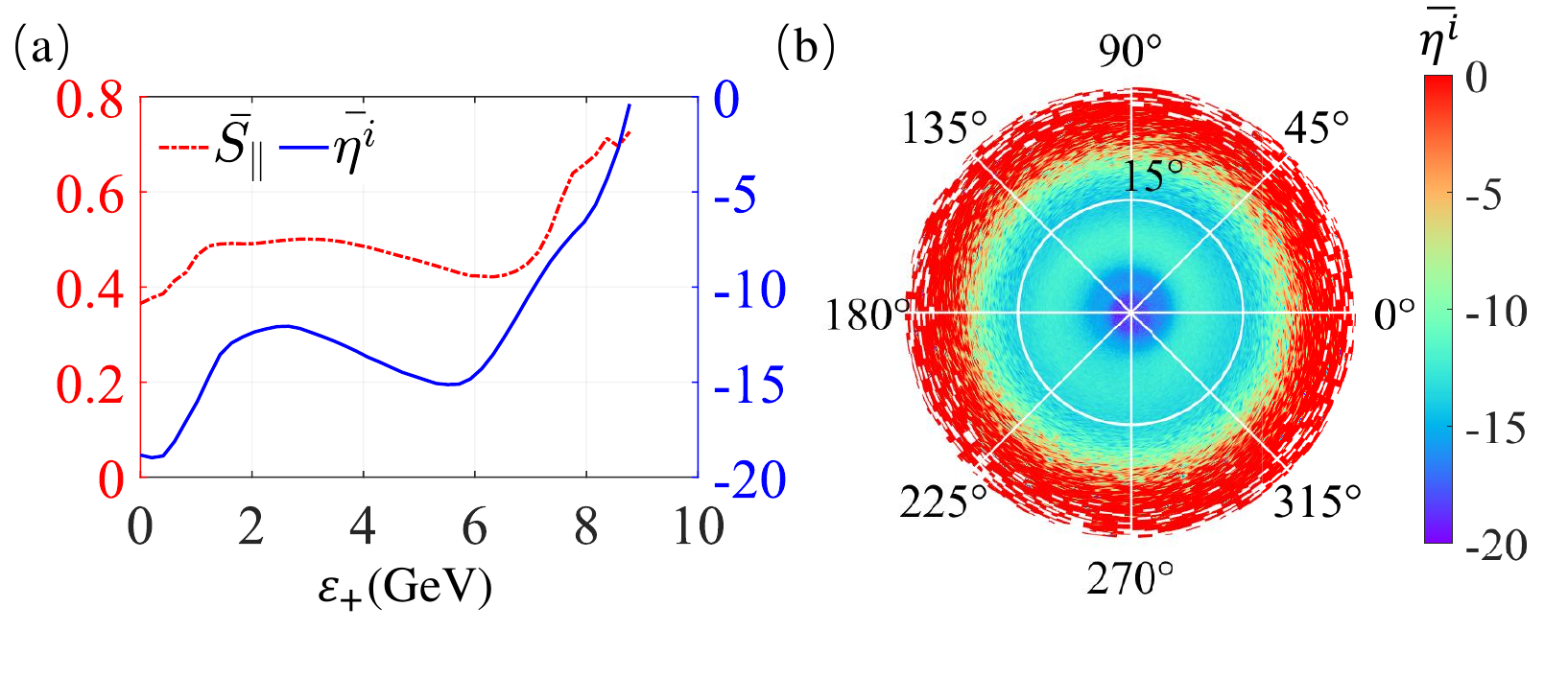}
         \label{}
    \end{minipage}
    \vskip -1.8 cc    
    \caption{\justifying  
    (a) Final longitudinal polarization $\bar{S}_{\parallel}$ (red dash-dotted) and creation phase $\bar{\eta^i}$ (blue solid) vs energy $\varepsilon_{+}$. (b) Birth phase distribution for positrons in Fig.~\ref{fig:2}(b).} 
      \label{fig:4}
      
    \vskip -0.3 cc
\end{figure}

\begin{figure}[b]
    \centering
    \begin{minipage}[b]{\linewidth}
        \centering
        \includegraphics[width=1.0\linewidth]{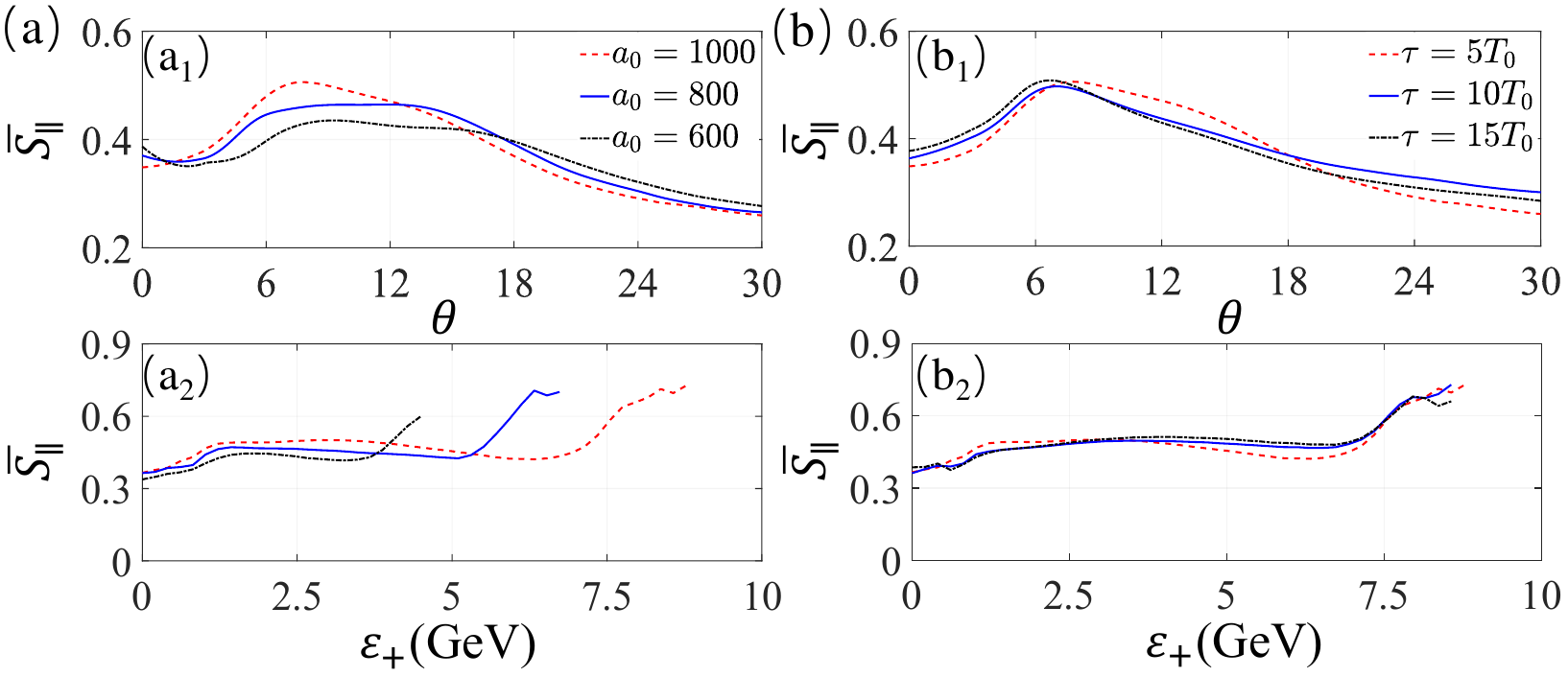}
 \label{}
    \end{minipage}
    \vskip -1.0 cc    
    \begin{minipage}[b]{\linewidth}
        \centering
        \includegraphics[width=1.0\linewidth]{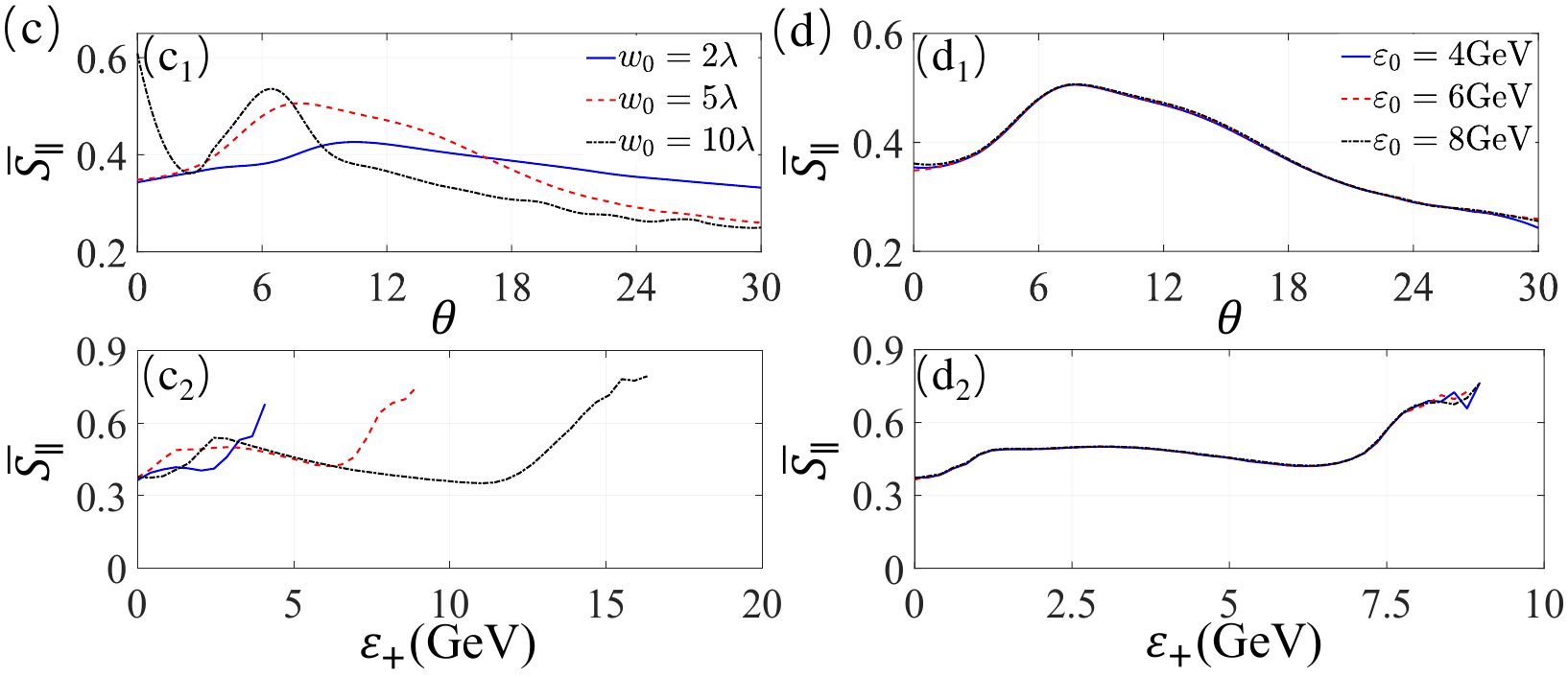}
         \label{}
    \end{minipage}
     \vskip -1.0 cc
     \caption{\justifying Profile curves of $\bar{S}_{\parallel}$ vs $\theta$ in $\phi=0^\circ$ and $\bar{S}_{\parallel}$ vs energy $\varepsilon_{+}$ in $\theta\in[0,12^\circ]$ with different laser intensity $a_0$ (a), laser pulse duration $\tau$ (b), laser focal radius $w_0$ (c) and initial seed electron energy $\varepsilon_0$ (d).  Other parameters are the same as those in Fig.\ref{fig:2}.}
    \label{fig:5}
     \vskip -0.3 cc
\end{figure}

The underlying mechanisms of positron polarization and acceleration are elucidated in Figs.\ref{fig3} and \ref{fig:4}. Parent photons exhibit a centrosymmetric radial polarization distribution, corresponding to a vanishing average polarization vector $\bar{\bm{\xi}} = (0,0,0)$ for the full beam (see Supplemental Material \cite{SM}), consistent with previous reports \cite{li2023highly}. Consequently, newly created positrons display negligible initial longitudinal polarization ($\overline{S_{\parallel}^i} = 1.2\%$) but substantial transverse polarization ($\overline{|S_{\perp}^i|} = 73\%$ \cite{SM}) [Fig.\ref{fig3}(a)] . These positrons are born with minimal transverse momenta and propagate primarily forward [Fig. \ref{fig3}(b)].

To unravel the governing dynamics, we simulate the trajectory and polarization evolution of a representative position in a plane-wave laser field with Gaussian temporal profile, neglecting radiation reaction (RR) [Fig.\ref{fig3}(c)]. The positron is initialized near the laser peak at phase $\eta'$ with energy 190 MeV (the mean energy for all the positrons at birth) and a polarization vector aligned with the instantaneous quantum spin axis \cite{li2020production, SM}—parallel to the electric field [$\bm{S} = (0,1,0)$]. Born within the laser field, the positron experiences an asymmetric field structure in the $y'$-direction [Fig.\ref{fig3}(c1)], acquiring a pronounced drift velocity of $p_{y'}$ [black solid curve, Fig.\ref{fig3}(c2)]. This leads to strong acceleration via the $\bm{v} \times \bf{B}$ mechanism [red dash-dotted curve, Fig.\ref{fig3}(c1)] along the laser propagation direction [red dash-dotted curve, Fig.\ref{fig3}(c2)], yielding significant final momentum of $p_z$. Throughout the interaction, spin precession dominates the polarization evolution: $S_\parallel$ rises from 0 to 0.396 while $S_\perp$ declines from 1 to 0.918, confirming efficient transverse-to-longitudinal polarization transfer [Fig.\ref{fig3}(c3)].

This conversion is described by the spin evolution equation \cite{jackson1998classical}:
\begin{equation}
\frac{dS_{\parallel}}{dt} = -\frac{e}{m}S_{\perp} \cdot \left[ \left( \frac{g}{2} - 1 \right) \boldsymbol{v} \times \mathbf{B} + \left( \frac{g v}{2} - \frac{1}{v} \right) \mathbf{E} \right],
\end{equation}
where $g$ is the energy-dependent gyromagnetic factor. For $g(\chi_e) = 2 + 2\mu(\chi_e)$ with $\mu(\chi_e) = \frac{\alpha}{\pi\chi_e}\int_0^\infty \frac{y}{(1+y)^3} {\bf L}{1/3}\left(\frac{2y}{3\chi_e}\right) dy$, we have $g \approx 2.00232$ when $\chi_e \ll 1$. The circularly polarized laser induces a phase shift in the spin precession—due to the anomalous magnetic moment ($g \ne 2$) —which is visible in the phase difference between 
$E_{y'}$ and $S_{y'}$ in Fig.\ref{fig3} (c). Although small, this shift is critical: it breaks the symmetry between spin and orbital motion, enabling efficient conversion from transverse to longitudinal polarization, as reflected in the synchronous evolution of $S_{\perp}$ and $S_{\parallel}$ in Fig.\ref{fig3} (c3)). The spin dynamics follow the precession equation, where the $\chi_e$ -dependence of the $g$-factor is essential. In our moderate-$\chi_e$ regime, $g$ slightly exceeds 2;  if $g$ were exactly 2, the spin would precess synchronously with the momentum, suppressing polarization conversion [see the $g=2$ case in Fig.\ref{fig3} (c3)].
Simultaneously, positrons gain energy via direct laser acceleration. The laser’s asymmetric electric field imparts a transverse drift along $y'$[see $p_{y'}$ in Fig. \ref{fig3} (c2)], and the $\bm{v} \times \bf{B}$ force rotates this motion efficiently along the propagation direction. The essential role of the $\bm{v} \times \bf{B}$ term is confirmed by control simulations where it is disabled, leading to enhanced $p_{y'}$ and negligible $p_{z'}$ [Fig.~\ref{fig3}(c2)].

Under realistic conditions with RR and a tightly focused laser beam [Fig.\ref{fig3}(d)], radiative energy losses damp the transverse momentum of the newborn positron, enabling its capture by the laser field [see Fig.\ref{fig3}(d1)]. This RR-mediated momentum reduction allows the positrons to be reflected and confined. The tightly focused geometry creates an asymmetric field that ensures sustained interaction before the positrons escape [see Fig.\ref{fig3}(d2)]. This capture-acceleration mechanism is distinctive in three key aspects: (i) tight integration with positron generation via QED cascades; (ii) constructive use of RR for particle capture, turning a typically detrimental effect into a beneficial tool; (iii) first demonstration of multi-GeV direct laser acceleration of positrons born within the pulse, with simultaneous longitudinal polarization via spin precession. Longer capture durations at higher energies further enhance the longitudinal polarization $S_\parallel$, which depends strongly on the birth laser phase [Fig.\ref{fig:4}(a)]. Significant radiation losses suppress kinetic energy, maintaining laser-field dominance over positron dynamics. Both the energy spectrum and angular distribution also exhibit strong phase dependence, as shown in Fig.\ref{fig:4}(b).


We systematically examine the parameter dependence of the average longitudinal polarization $\bar{S}_{\parallel}$ in Fig.\ref{fig:5}. Increasing laser intensity ($a_0$ from 600 to 1000) enhances both positron energy and polarization due to stronger electromagnetic fields [Fig.\ref{fig:5}(a)], with corresponding yields of 14.6, 20.5, 25.8 $e^+/e^-$ respectively. The polarization shows weak dependence on the pulse duration over the range $5T_0$ to $15T_0$ [Fig.\ref{fig:5}(b)], consistent with the fact that the positron's final state is governed by the peak laser intensity \cite{SM}. Varying the laser focal radius ($w_0$ from $2\lambda_0$ to $10\lambda_0$) alters the polarization distribution, as it modifies the field phase sampled by the positrons; a larger radius also promotes higher acceleration and energy gain [Fig.\ref{fig:5}(c)]. Changes in the seed electron energy ($\varepsilon_0$) have a negligible effect on polarization [Fig.\ref{fig:5}(d)]: radiation-dominated reflection dynamics render the process insensitive to initial energy. However, $\varepsilon_0$ strongly influences the positron yield via the parent photon spectrum, increasing from 17.0 to 25.8 to 34.7 $e^+/e^-$ at $\varepsilon_0$ = 4, 6, and 8 GeV respectively.

In conclusion, we have demonstrated a compact all-optical scheme that simultaneously resolves the long-standing challenges of yield, energy, and polarization in laser-driven positron beam generation. Our approach integrates positron production via a QED cascade, multi-GeV acceleration, and spin manipulation into a single laser-electron collision stage. By leveraging radiation reaction to capture newborn positrons within the laser field, direct laser acceleration boosts them to GeV energies while intrinsic spin precession rotates their polarization to the longitudinal direction. The resulting beams achieve unprecedented performance: a high yield of 25.8 $e^+/e^-$, energies up to 9 GeV, and longitudinal polarization exceeding 70\%.  With a feasible seed electron beam density of  $10^{8}$ per bunch \cite{gonsalves2019petawatt}, this setup produces $10^{9}$ positrons per shot. Employing ultrahigh-charge ($\sim$ 100 nC) electron beams \cite{Ma2018} further enables positron densities of $\sim 10^{12}$ per bunch,  meeting the requirements of future colliders and fundamental physics experiments. This work thereby establishes a promising platform for highly polarized positron sources applicable to studies of chiral dynamics and new physics beyond the Standard Model.

\section{Acknowledgements} This work is supported by the National Natural Science
Foundation of China (Grants nos.12222507 and 12075187), the Strategic
Priority Research Program of the Chinese Academy of Sciences (Grants nos.
XDA25031000 and No.XDA25010300).

\nocite{*}
\bibliography{reference}

\begin{thebibliography}{78}%
\makeatletter
\providecommand \@ifxundefined [1]{%
 \@ifx{#1\undefined}
}%
\providecommand \@ifnum [1]{%
 \ifnum #1\expandafter \@firstoftwo
 \else \expandafter \@secondoftwo
 \fi
}%
\providecommand \@ifx [1]{%
 \ifx #1\expandafter \@firstoftwo
 \else \expandafter \@secondoftwo
 \fi
}%
\providecommand \natexlab [1]{#1}%
\providecommand \enquote  [1]{``#1''}%
\providecommand \bibnamefont  [1]{#1}%
\providecommand \bibfnamefont [1]{#1}%
\providecommand \citenamefont [1]{#1}%
\providecommand \href@noop [0]{\@secondoftwo}%
\providecommand \href [0]{\begingroup \@sanitize@url \@href}%
\providecommand \@href[1]{\@@startlink{#1}\@@href}%
\providecommand \@@href[1]{\endgroup#1\@@endlink}%
\providecommand \@sanitize@url [0]{\catcode `\\12\catcode `\$12\catcode `\&12\catcode `\#12\catcode `\^12\catcode `\_12\catcode `\%12\relax}%
\providecommand \@@startlink[1]{}%
\providecommand \@@endlink[0]{}%
\providecommand \url  [0]{\begingroup\@sanitize@url \@url }%
\providecommand \@url [1]{\endgroup\@href {#1}{\urlprefix }}%
\providecommand \urlprefix  [0]{URL }%
\providecommand \Eprint [0]{\href }%
\providecommand \doibase [0]{https://doi.org/}%
\providecommand \selectlanguage [0]{\@gobble}%
\providecommand \bibinfo  [0]{\@secondoftwo}%
\providecommand \bibfield  [0]{\@secondoftwo}%
\providecommand \translation [1]{[#1]}%
\providecommand \BibitemOpen [0]{}%
\providecommand \bibitemStop [0]{}%
\providecommand \bibitemNoStop [0]{.\EOS\space}%
\providecommand \EOS [0]{\spacefactor3000\relax}%
\providecommand \BibitemShut  [1]{\csname bibitem#1\endcsname}%
\let\auto@bib@innerbib\@empty
\bibitem [{\citenamefont {Bambade}\ \emph {et~al.}(2019)\citenamefont {Bambade}, \citenamefont {Barklow}, \citenamefont {Behnke}, \citenamefont {Berggren}, \citenamefont {Brau}, \citenamefont {Burrows}, \citenamefont {Denisov}, \citenamefont {Faus-Golfe}, \citenamefont {Foster}, \citenamefont {Fujii} \emph {et~al.}}]{bambade2019international}%
  \BibitemOpen
  \bibfield  {author} {\bibinfo {author} {\bibfnamefont {P.}~\bibnamefont {Bambade}}, \bibinfo {author} {\bibfnamefont {T.}~\bibnamefont {Barklow}}, \bibinfo {author} {\bibfnamefont {T.}~\bibnamefont {Behnke}}, \bibinfo {author} {\bibfnamefont {M.}~\bibnamefont {Berggren}}, \bibinfo {author} {\bibfnamefont {J.}~\bibnamefont {Brau}}, \bibinfo {author} {\bibfnamefont {P.}~\bibnamefont {Burrows}}, \bibinfo {author} {\bibfnamefont {D.}~\bibnamefont {Denisov}}, \bibinfo {author} {\bibfnamefont {A.}~\bibnamefont {Faus-Golfe}}, \bibinfo {author} {\bibfnamefont {B.}~\bibnamefont {Foster}}, \bibinfo {author} {\bibfnamefont {K.}~\bibnamefont {Fujii}}, \emph {et~al.},\ }\bibfield  {title} {\bibinfo {title} {The international linear collider: A global project},\ }\href@noop {} {\bibfield  {journal} {\bibinfo  {journal} {arXiv preprint arXiv:1903.01629}\ } (\bibinfo {year} {2019})}\BibitemShut {NoStop}%
\bibitem [{\citenamefont {Kovalenko}\ \emph {et~al.}(2012)\citenamefont {Kovalenko}, \citenamefont {Adeyemi}, \citenamefont {Hartin}, \citenamefont {Moortgat-Pick}, \citenamefont {Malysheva}, \citenamefont {Riemann}, \citenamefont {Staufenbiel},\ and\ \citenamefont {Ushakov}}]{Kovalenko2012SpinTracking}%
  \BibitemOpen
  \bibfield  {author} {\bibinfo {author} {\bibfnamefont {V.}~\bibnamefont {Kovalenko}}, \bibinfo {author} {\bibfnamefont {O.~S.}\ \bibnamefont {Adeyemi}}, \bibinfo {author} {\bibfnamefont {A.}~\bibnamefont {Hartin}}, \bibinfo {author} {\bibfnamefont {G.}~\bibnamefont {Moortgat-Pick}}, \bibinfo {author} {\bibfnamefont {L.}~\bibnamefont {Malysheva}}, \bibinfo {author} {\bibfnamefont {S.}~\bibnamefont {Riemann}}, \bibinfo {author} {\bibfnamefont {F.}~\bibnamefont {Staufenbiel}},\ and\ \bibinfo {author} {\bibfnamefont {A.}~\bibnamefont {Ushakov}},\ }\bibfield  {title} {\bibinfo {title} {Spin tracking at the ilc positron source},\ }\href@noop {} {\bibfield  {journal} {\bibinfo  {journal} {arXiv preprint}\ } (\bibinfo {year} {2012})},\ \bibinfo {note} {presented at POSIPOL 2011 Workshop},\ \Eprint {https://arxiv.org/abs/1202.0751} {arXiv:1202.0751 [physics.acc-ph]} \BibitemShut {NoStop}%
\bibitem [{\citenamefont {Seipt}\ \emph {et~al.}(2021)\citenamefont {Seipt}, \citenamefont {Ridgers}, \citenamefont {Sorbo},\ and\ \citenamefont {Thomas}}]{Seipt2021PolarizedQEDCascades}%
  \BibitemOpen
  \bibfield  {author} {\bibinfo {author} {\bibfnamefont {D.}~\bibnamefont {Seipt}}, \bibinfo {author} {\bibfnamefont {C.~P.}\ \bibnamefont {Ridgers}}, \bibinfo {author} {\bibfnamefont {D.~D.}\ \bibnamefont {Sorbo}},\ and\ \bibinfo {author} {\bibfnamefont {A.~G.~R.}\ \bibnamefont {Thomas}},\ }\bibfield  {title} {\bibinfo {title} {Polarized {QED} cascades},\ }\href {https://doi.org/10.1088/1367-2630/abf584} {\bibfield  {journal} {\bibinfo  {journal} {New Journal of Physics}\ }\textbf {\bibinfo {volume} {23}},\ \bibinfo {pages} {053025} (\bibinfo {year} {2021})},\ \Eprint {https://arxiv.org/abs/2010.04078} {arXiv:2010.04078 [physics.plasm-ph]} \BibitemShut {NoStop}%
\bibitem [{\citenamefont {Moortgat-Pick}\ \emph {et~al.}(2008)\citenamefont {Moortgat-Pick}, \citenamefont {Abe}, \citenamefont {Alexander}, \citenamefont {Ananthanarayan}, \citenamefont {Babich}, \citenamefont {Bharadwaj}, \citenamefont {Barber}, \citenamefont {Bartl}, \citenamefont {Brachmann}, \citenamefont {Chen} \emph {et~al.}}]{moortgat2008polarized}%
  \BibitemOpen
  \bibfield  {author} {\bibinfo {author} {\bibfnamefont {G.}~\bibnamefont {Moortgat-Pick}}, \bibinfo {author} {\bibfnamefont {T.}~\bibnamefont {Abe}}, \bibinfo {author} {\bibfnamefont {G.}~\bibnamefont {Alexander}}, \bibinfo {author} {\bibfnamefont {B.}~\bibnamefont {Ananthanarayan}}, \bibinfo {author} {\bibfnamefont {A.}~\bibnamefont {Babich}}, \bibinfo {author} {\bibfnamefont {V.}~\bibnamefont {Bharadwaj}}, \bibinfo {author} {\bibfnamefont {D.}~\bibnamefont {Barber}}, \bibinfo {author} {\bibfnamefont {A.}~\bibnamefont {Bartl}}, \bibinfo {author} {\bibfnamefont {A.}~\bibnamefont {Brachmann}}, \bibinfo {author} {\bibfnamefont {S.}~\bibnamefont {Chen}}, \emph {et~al.},\ }\bibfield  {title} {\bibinfo {title} {Polarized positrons and electrons at the linear collider},\ }\href@noop {} {\bibfield  {journal} {\bibinfo  {journal} {Physics Reports}\ }\textbf {\bibinfo {volume} {460}},\ \bibinfo {pages} {131} (\bibinfo {year} {2008})}\BibitemShut {NoStop}%
\bibitem [{\citenamefont {Richard}\ \emph {et~al.}(2001)\citenamefont {Richard}, \citenamefont {Schneider}, \citenamefont {Trines},\ and\ \citenamefont {Wagner}}]{richard2001tesla}%
  \BibitemOpen
  \bibfield  {author} {\bibinfo {author} {\bibfnamefont {F.}~\bibnamefont {Richard}}, \bibinfo {author} {\bibfnamefont {J.}~\bibnamefont {Schneider}}, \bibinfo {author} {\bibfnamefont {D.}~\bibnamefont {Trines}},\ and\ \bibinfo {author} {\bibfnamefont {A.}~\bibnamefont {Wagner}},\ }\bibfield  {title} {\bibinfo {title} {Tesla technical design report part i: Executive summary},\ }\href@noop {} {\bibfield  {journal} {\bibinfo  {journal} {arXiv preprint hep-ph/0106314}\ } (\bibinfo {year} {2001})}\BibitemShut {NoStop}%
\bibitem [{\citenamefont {Zitzewitz}\ \emph {et~al.}(1979)\citenamefont {Zitzewitz}, \citenamefont {Van~House}, \citenamefont {Rich},\ and\ \citenamefont {Gidley}}]{zitzewitz1979spin}%
  \BibitemOpen
  \bibfield  {author} {\bibinfo {author} {\bibfnamefont {P.}~\bibnamefont {Zitzewitz}}, \bibinfo {author} {\bibfnamefont {J.}~\bibnamefont {Van~House}}, \bibinfo {author} {\bibfnamefont {A.}~\bibnamefont {Rich}},\ and\ \bibinfo {author} {\bibfnamefont {D.}~\bibnamefont {Gidley}},\ }\bibfield  {title} {\bibinfo {title} {Spin polarization of low-energy positron beams},\ }\href@noop {} {\bibfield  {journal} {\bibinfo  {journal} {Physical Review Letters}\ }\textbf {\bibinfo {volume} {43}},\ \bibinfo {pages} {1281} (\bibinfo {year} {1979})}\BibitemShut {NoStop}%
\bibitem [{\citenamefont {Sokolov}\ and\ \citenamefont {Ternov}(1964)}]{sokolov1964polarization}%
  \BibitemOpen
  \bibfield  {author} {\bibinfo {author} {\bibfnamefont {A.}~\bibnamefont {Sokolov}}\ and\ \bibinfo {author} {\bibfnamefont {M.}~\bibnamefont {Ternov}},\ }\bibfield  {title} {\bibinfo {title} {On polarization and spin effects in the theory of synchrotron radiation},\ }in\ \href@noop {} {\emph {\bibinfo {booktitle} {Sov. Phys.-Dokl.}}},\ Vol.~\bibinfo {volume} {8}\ (\bibinfo {year} {1964})\ pp.\ \bibinfo {pages} {1203--1205}\BibitemShut {NoStop}%
\bibitem [{\citenamefont {Potylitsin}(1997)}]{potylitsin1997production}%
  \BibitemOpen
  \bibfield  {author} {\bibinfo {author} {\bibfnamefont {A.}~\bibnamefont {Potylitsin}},\ }\bibfield  {title} {\bibinfo {title} {Production of polarized positrons through interaction of longitudinally polarized electrons with thin targets},\ }\href@noop {} {\bibfield  {journal} {\bibinfo  {journal} {Nuclear Instruments and Methods in Physics Research Section A: Accelerators, Spectrometers, Detectors and Associated Equipment}\ }\textbf {\bibinfo {volume} {398}},\ \bibinfo {pages} {395} (\bibinfo {year} {1997})}\BibitemShut {NoStop}%
\bibitem [{\citenamefont {Olsen}\ and\ \citenamefont {Maximon}(1959)}]{olsen1959photon}%
  \BibitemOpen
  \bibfield  {author} {\bibinfo {author} {\bibfnamefont {H.}~\bibnamefont {Olsen}}\ and\ \bibinfo {author} {\bibfnamefont {L.}~\bibnamefont {Maximon}},\ }\bibfield  {title} {\bibinfo {title} {Photon and electron polarization in high-energy bremsstrahlung and pair production with screening},\ }\href@noop {} {\bibfield  {journal} {\bibinfo  {journal} {Physical Review}\ }\textbf {\bibinfo {volume} {114}},\ \bibinfo {pages} {887} (\bibinfo {year} {1959})}\BibitemShut {NoStop}%
\bibitem [{\citenamefont {Abbott}\ \emph {et~al.}(2016{\natexlab{a}})\citenamefont {Abbott}, \citenamefont {Adderley}, \citenamefont {Adeyemi}, \citenamefont {Aguilera}, \citenamefont {Ali}, \citenamefont {Areti}, \citenamefont {Baylac}, \citenamefont {Benesch}, \citenamefont {Bosson}, \citenamefont {Cade}, \citenamefont {Camsonne}, \citenamefont {Cardman}, \citenamefont {Clark}, \citenamefont {Cole}, \citenamefont {Covert}, \citenamefont {Cuevas}, \citenamefont {Dadoun}, \citenamefont {Dale}, \citenamefont {Dong}, \citenamefont {Dumas}, \citenamefont {Fanchini}, \citenamefont {Forest}, \citenamefont {Forman}, \citenamefont {Freyberger}, \citenamefont {Froidefond}, \citenamefont {Golge}, \citenamefont {Grames}, \citenamefont {Gu\`eye}, \citenamefont {Hansknecht}, \citenamefont {Harrell}, \citenamefont {Hoskins}, \citenamefont {Hyde}, \citenamefont {Josey}, \citenamefont {Kazimi}, \citenamefont {Kim}, \citenamefont {Machie}, \citenamefont {Mahoney}, \citenamefont {Mammei}, \citenamefont {Marton}, \citenamefont
  {McCarter}, \citenamefont {McCaughan}, \citenamefont {McHugh}, \citenamefont {McNulty}, \citenamefont {Mesick}, \citenamefont {Michaelides}, \citenamefont {Michaels}, \citenamefont {Moffit}, \citenamefont {Moser}, \citenamefont {Mu\~noz Camacho}, \citenamefont {Muraz}, \citenamefont {Opper}, \citenamefont {Poelker}, \citenamefont {R\'eal}, \citenamefont {Richardson}, \citenamefont {Setiniyaz}, \citenamefont {Stutzman}, \citenamefont {Suleiman}, \citenamefont {Tennant}, \citenamefont {Tsai}, \citenamefont {Turner}, \citenamefont {Ungaro}, \citenamefont {Variola}, \citenamefont {Voutier}, \citenamefont {Wang},\ and\ \citenamefont {Zhang}}]{Abbott2016}%
  \BibitemOpen
  \bibfield  {author} {\bibinfo {author} {\bibfnamefont {D.}~\bibnamefont {Abbott}}, \bibinfo {author} {\bibfnamefont {P.}~\bibnamefont {Adderley}}, \bibinfo {author} {\bibfnamefont {A.}~\bibnamefont {Adeyemi}}, \bibinfo {author} {\bibfnamefont {P.}~\bibnamefont {Aguilera}}, \bibinfo {author} {\bibfnamefont {M.}~\bibnamefont {Ali}}, \bibinfo {author} {\bibfnamefont {H.}~\bibnamefont {Areti}}, \bibinfo {author} {\bibfnamefont {M.}~\bibnamefont {Baylac}}, \bibinfo {author} {\bibfnamefont {J.}~\bibnamefont {Benesch}}, \bibinfo {author} {\bibfnamefont {G.}~\bibnamefont {Bosson}}, \bibinfo {author} {\bibfnamefont {B.}~\bibnamefont {Cade}}, \bibinfo {author} {\bibfnamefont {A.}~\bibnamefont {Camsonne}}, \bibinfo {author} {\bibfnamefont {L.~S.}\ \bibnamefont {Cardman}}, \bibinfo {author} {\bibfnamefont {J.}~\bibnamefont {Clark}}, \bibinfo {author} {\bibfnamefont {P.}~\bibnamefont {Cole}}, \bibinfo {author} {\bibfnamefont {S.}~\bibnamefont {Covert}}, \bibinfo {author} {\bibfnamefont {C.}~\bibnamefont {Cuevas}},
  \bibinfo {author} {\bibfnamefont {O.}~\bibnamefont {Dadoun}}, \bibinfo {author} {\bibfnamefont {D.}~\bibnamefont {Dale}}, \bibinfo {author} {\bibfnamefont {H.}~\bibnamefont {Dong}}, \bibinfo {author} {\bibfnamefont {J.}~\bibnamefont {Dumas}}, \bibinfo {author} {\bibfnamefont {E.}~\bibnamefont {Fanchini}}, \bibinfo {author} {\bibfnamefont {T.}~\bibnamefont {Forest}}, \bibinfo {author} {\bibfnamefont {E.}~\bibnamefont {Forman}}, \bibinfo {author} {\bibfnamefont {A.}~\bibnamefont {Freyberger}}, \bibinfo {author} {\bibfnamefont {E.}~\bibnamefont {Froidefond}}, \bibinfo {author} {\bibfnamefont {S.}~\bibnamefont {Golge}}, \bibinfo {author} {\bibfnamefont {J.}~\bibnamefont {Grames}}, \bibinfo {author} {\bibfnamefont {P.}~\bibnamefont {Gu\`eye}}, \bibinfo {author} {\bibfnamefont {J.}~\bibnamefont {Hansknecht}}, \bibinfo {author} {\bibfnamefont {P.}~\bibnamefont {Harrell}}, \bibinfo {author} {\bibfnamefont {J.}~\bibnamefont {Hoskins}}, \bibinfo {author} {\bibfnamefont {C.}~\bibnamefont {Hyde}}, \bibinfo {author}
  {\bibfnamefont {B.}~\bibnamefont {Josey}}, \bibinfo {author} {\bibfnamefont {R.}~\bibnamefont {Kazimi}}, \bibinfo {author} {\bibfnamefont {Y.}~\bibnamefont {Kim}}, \bibinfo {author} {\bibfnamefont {D.}~\bibnamefont {Machie}}, \bibinfo {author} {\bibfnamefont {K.}~\bibnamefont {Mahoney}}, \bibinfo {author} {\bibfnamefont {R.}~\bibnamefont {Mammei}}, \bibinfo {author} {\bibfnamefont {M.}~\bibnamefont {Marton}}, \bibinfo {author} {\bibfnamefont {J.}~\bibnamefont {McCarter}}, \bibinfo {author} {\bibfnamefont {M.}~\bibnamefont {McCaughan}}, \bibinfo {author} {\bibfnamefont {M.}~\bibnamefont {McHugh}}, \bibinfo {author} {\bibfnamefont {D.}~\bibnamefont {McNulty}}, \bibinfo {author} {\bibfnamefont {K.~E.}\ \bibnamefont {Mesick}}, \bibinfo {author} {\bibfnamefont {T.}~\bibnamefont {Michaelides}}, \bibinfo {author} {\bibfnamefont {R.}~\bibnamefont {Michaels}}, \bibinfo {author} {\bibfnamefont {B.}~\bibnamefont {Moffit}}, \bibinfo {author} {\bibfnamefont {D.}~\bibnamefont {Moser}}, \bibinfo {author} {\bibfnamefont
  {C.}~\bibnamefont {Mu\~noz Camacho}}, \bibinfo {author} {\bibfnamefont {J.-F.}\ \bibnamefont {Muraz}}, \bibinfo {author} {\bibfnamefont {A.}~\bibnamefont {Opper}}, \bibinfo {author} {\bibfnamefont {M.}~\bibnamefont {Poelker}}, \bibinfo {author} {\bibfnamefont {J.-S.}\ \bibnamefont {R\'eal}}, \bibinfo {author} {\bibfnamefont {L.}~\bibnamefont {Richardson}}, \bibinfo {author} {\bibfnamefont {S.}~\bibnamefont {Setiniyaz}}, \bibinfo {author} {\bibfnamefont {M.}~\bibnamefont {Stutzman}}, \bibinfo {author} {\bibfnamefont {R.}~\bibnamefont {Suleiman}}, \bibinfo {author} {\bibfnamefont {C.}~\bibnamefont {Tennant}}, \bibinfo {author} {\bibfnamefont {C.}~\bibnamefont {Tsai}}, \bibinfo {author} {\bibfnamefont {D.}~\bibnamefont {Turner}}, \bibinfo {author} {\bibfnamefont {M.}~\bibnamefont {Ungaro}}, \bibinfo {author} {\bibfnamefont {A.}~\bibnamefont {Variola}}, \bibinfo {author} {\bibfnamefont {E.}~\bibnamefont {Voutier}}, \bibinfo {author} {\bibfnamefont {Y.}~\bibnamefont {Wang}},\ and\ \bibinfo {author}
  {\bibfnamefont {Y.}~\bibnamefont {Zhang}} (\bibinfo {collaboration} {PEPPo Collaboration}),\ }\bibfield  {title} {\bibinfo {title} {Production of highly polarized positrons using polarized electrons at mev energies},\ }\href {https://doi.org/10.1103/PhysRevLett.116.214801} {\bibfield  {journal} {\bibinfo  {journal} {Phys. Rev. Lett.}\ }\textbf {\bibinfo {volume} {116}},\ \bibinfo {pages} {214801} (\bibinfo {year} {2016}{\natexlab{a}})}\BibitemShut {NoStop}%
\bibitem [{\citenamefont {Omori}\ \emph {et~al.}(2006)\citenamefont {Omori}, \citenamefont {Fukuda}, \citenamefont {Hirose}, \citenamefont {Kurihara}, \citenamefont {Kuroda}, \citenamefont {Nomura}, \citenamefont {Ohashi}, \citenamefont {Okugi}, \citenamefont {Sakaue}, \citenamefont {Saito}, \citenamefont {Urakawa}, \citenamefont {Washio},\ and\ \citenamefont {Yamazaki}}]{Omori2006}%
  \BibitemOpen
  \bibfield  {author} {\bibinfo {author} {\bibfnamefont {T.}~\bibnamefont {Omori}}, \bibinfo {author} {\bibfnamefont {M.}~\bibnamefont {Fukuda}}, \bibinfo {author} {\bibfnamefont {T.}~\bibnamefont {Hirose}}, \bibinfo {author} {\bibfnamefont {Y.}~\bibnamefont {Kurihara}}, \bibinfo {author} {\bibfnamefont {R.}~\bibnamefont {Kuroda}}, \bibinfo {author} {\bibfnamefont {M.}~\bibnamefont {Nomura}}, \bibinfo {author} {\bibfnamefont {A.}~\bibnamefont {Ohashi}}, \bibinfo {author} {\bibfnamefont {T.}~\bibnamefont {Okugi}}, \bibinfo {author} {\bibfnamefont {K.}~\bibnamefont {Sakaue}}, \bibinfo {author} {\bibfnamefont {T.}~\bibnamefont {Saito}}, \bibinfo {author} {\bibfnamefont {J.}~\bibnamefont {Urakawa}}, \bibinfo {author} {\bibfnamefont {M.}~\bibnamefont {Washio}},\ and\ \bibinfo {author} {\bibfnamefont {I.}~\bibnamefont {Yamazaki}},\ }\bibfield  {title} {\bibinfo {title} {Efficient propagation of polarization from laser photons to positrons through compton scattering and electron-positron pair creation},\ }\href
  {https://doi.org/10.1103/PhysRevLett.96.114801} {\bibfield  {journal} {\bibinfo  {journal} {Phys. Rev. Lett.}\ }\textbf {\bibinfo {volume} {96}},\ \bibinfo {pages} {114801} (\bibinfo {year} {2006})}\BibitemShut {NoStop}%
\bibitem [{\citenamefont {Chen}\ \emph {et~al.}(2019)\citenamefont {Chen}, \citenamefont {He}, \citenamefont {Shaisultanov}, \citenamefont {Hatsagortsyan},\ and\ \citenamefont {Keitel}}]{chen2019polarized}%
  \BibitemOpen
  \bibfield  {author} {\bibinfo {author} {\bibfnamefont {Y.-Y.}\ \bibnamefont {Chen}}, \bibinfo {author} {\bibfnamefont {P.-L.}\ \bibnamefont {He}}, \bibinfo {author} {\bibfnamefont {R.}~\bibnamefont {Shaisultanov}}, \bibinfo {author} {\bibfnamefont {K.~Z.}\ \bibnamefont {Hatsagortsyan}},\ and\ \bibinfo {author} {\bibfnamefont {C.~H.}\ \bibnamefont {Keitel}},\ }\bibfield  {title} {\bibinfo {title} {Polarized positron beams via intense two-color laser pulses},\ }\href@noop {} {\bibfield  {journal} {\bibinfo  {journal} {Physical review letters}\ }\textbf {\bibinfo {volume} {123}},\ \bibinfo {pages} {174801} (\bibinfo {year} {2019})}\BibitemShut {NoStop}%
\bibitem [{\citenamefont {Wan}\ \emph {et~al.}(2020)\citenamefont {Wan}, \citenamefont {Shaisultanov}, \citenamefont {Li}, \citenamefont {Hatsagortsyan}, \citenamefont {Keitel},\ and\ \citenamefont {Li}}]{wan2020ultrarelativistic}%
  \BibitemOpen
  \bibfield  {author} {\bibinfo {author} {\bibfnamefont {F.}~\bibnamefont {Wan}}, \bibinfo {author} {\bibfnamefont {R.}~\bibnamefont {Shaisultanov}}, \bibinfo {author} {\bibfnamefont {Y.-F.}\ \bibnamefont {Li}}, \bibinfo {author} {\bibfnamefont {K.~Z.}\ \bibnamefont {Hatsagortsyan}}, \bibinfo {author} {\bibfnamefont {C.~H.}\ \bibnamefont {Keitel}},\ and\ \bibinfo {author} {\bibfnamefont {J.-X.}\ \bibnamefont {Li}},\ }\bibfield  {title} {\bibinfo {title} {Ultrarelativistic polarized positron jets via collision of electron and ultraintense laser beams},\ }\href@noop {} {\bibfield  {journal} {\bibinfo  {journal} {Physics Letters B}\ }\textbf {\bibinfo {volume} {800}},\ \bibinfo {pages} {135120} (\bibinfo {year} {2020})}\BibitemShut {NoStop}%
\bibitem [{\citenamefont {Zhuang}\ \emph {et~al.}(2023)\citenamefont {Zhuang}, \citenamefont {Chen}, \citenamefont {Li}, \citenamefont {Hatsagortsyan},\ and\ \citenamefont {Keitel}}]{zhuang2023laser}%
  \BibitemOpen
  \bibfield  {author} {\bibinfo {author} {\bibfnamefont {K.-H.}\ \bibnamefont {Zhuang}}, \bibinfo {author} {\bibfnamefont {Y.-Y.}\ \bibnamefont {Chen}}, \bibinfo {author} {\bibfnamefont {Y.-F.}\ \bibnamefont {Li}}, \bibinfo {author} {\bibfnamefont {K.~Z.}\ \bibnamefont {Hatsagortsyan}},\ and\ \bibinfo {author} {\bibfnamefont {C.~H.}\ \bibnamefont {Keitel}},\ }\bibfield  {title} {\bibinfo {title} {Laser-driven lepton polarization in the quantum radiation-dominated reflection regime},\ }\href@noop {} {\bibfield  {journal} {\bibinfo  {journal} {Physical Review D}\ }\textbf {\bibinfo {volume} {108}},\ \bibinfo {pages} {033001} (\bibinfo {year} {2023})}\BibitemShut {NoStop}%
\bibitem [{\citenamefont {Xue}\ \emph {et~al.}(2023)\citenamefont {Xue}, \citenamefont {Sun}, \citenamefont {Wei}, \citenamefont {Li}, \citenamefont {Zhao}, \citenamefont {Wan}, \citenamefont {Lv}, \citenamefont {Zhao}, \citenamefont {Xu},\ and\ \citenamefont {Li}}]{xue2023generation}%
  \BibitemOpen
  \bibfield  {author} {\bibinfo {author} {\bibfnamefont {K.}~\bibnamefont {Xue}}, \bibinfo {author} {\bibfnamefont {T.}~\bibnamefont {Sun}}, \bibinfo {author} {\bibfnamefont {K.-J.}\ \bibnamefont {Wei}}, \bibinfo {author} {\bibfnamefont {Z.-P.}\ \bibnamefont {Li}}, \bibinfo {author} {\bibfnamefont {Q.}~\bibnamefont {Zhao}}, \bibinfo {author} {\bibfnamefont {F.}~\bibnamefont {Wan}}, \bibinfo {author} {\bibfnamefont {C.}~\bibnamefont {Lv}}, \bibinfo {author} {\bibfnamefont {Y.-T.}\ \bibnamefont {Zhao}}, \bibinfo {author} {\bibfnamefont {Z.-F.}\ \bibnamefont {Xu}},\ and\ \bibinfo {author} {\bibfnamefont {J.-X.}\ \bibnamefont {Li}},\ }\bibfield  {title} {\bibinfo {title} {Generation of high-density high-polarization positrons via single-shot strong laser-foil interaction},\ }\href@noop {} {\bibfield  {journal} {\bibinfo  {journal} {Physical Review Letters}\ }\textbf {\bibinfo {volume} {131}},\ \bibinfo {pages} {175101} (\bibinfo {year} {2023})}\BibitemShut {NoStop}%
\bibitem [{\citenamefont {Song}\ \emph {et~al.}(2022)\citenamefont {Song}, \citenamefont {Wang},\ and\ \citenamefont {Li}}]{song2022dense}%
  \BibitemOpen
  \bibfield  {author} {\bibinfo {author} {\bibfnamefont {H.-H.}\ \bibnamefont {Song}}, \bibinfo {author} {\bibfnamefont {W.-M.}\ \bibnamefont {Wang}},\ and\ \bibinfo {author} {\bibfnamefont {Y.-T.}\ \bibnamefont {Li}},\ }\bibfield  {title} {\bibinfo {title} {Dense polarized positrons from laser-irradiated foil targets in the qed regime},\ }\href@noop {} {\bibfield  {journal} {\bibinfo  {journal} {Physical Review Letters}\ }\textbf {\bibinfo {volume} {129}},\ \bibinfo {pages} {035001} (\bibinfo {year} {2022})}\BibitemShut {NoStop}%
\bibitem [{\citenamefont {Zhu}\ \emph {et~al.}(2024)\citenamefont {Zhu}, \citenamefont {Liu}, \citenamefont {Yu}, \citenamefont {Chen}, \citenamefont {Weng}, \citenamefont {Wang},\ and\ \citenamefont {Sheng}}]{zhu2024dense}%
  \BibitemOpen
  \bibfield  {author} {\bibinfo {author} {\bibfnamefont {X.-L.}\ \bibnamefont {Zhu}}, \bibinfo {author} {\bibfnamefont {W.-Y.}\ \bibnamefont {Liu}}, \bibinfo {author} {\bibfnamefont {T.-P.}\ \bibnamefont {Yu}}, \bibinfo {author} {\bibfnamefont {M.}~\bibnamefont {Chen}}, \bibinfo {author} {\bibfnamefont {S.-M.}\ \bibnamefont {Weng}}, \bibinfo {author} {\bibfnamefont {W.-M.}\ \bibnamefont {Wang}},\ and\ \bibinfo {author} {\bibfnamefont {Z.-M.}\ \bibnamefont {Sheng}},\ }\bibfield  {title} {\bibinfo {title} {Dense polarized positrons from beam-solid interaction},\ }\href@noop {} {\bibfield  {journal} {\bibinfo  {journal} {Physical Review Letters}\ }\textbf {\bibinfo {volume} {132}},\ \bibinfo {pages} {235001} (\bibinfo {year} {2024})}\BibitemShut {NoStop}%
\bibitem [{\citenamefont {Flottmann}(1993)}]{Flottmann1993}%
  \BibitemOpen
  \bibfield  {author} {\bibinfo {author} {\bibfnamefont {K.}~\bibnamefont {Flottmann}},\ }\emph {\bibinfo {title} {{Investigations toward the development of polarized and unpolarized high intensity positron sources for linear colliders}}},\ \href@noop {} {\bibinfo {type} {Other thesis}} (\bibinfo {year} {1993})\BibitemShut {NoStop}%
\bibitem [{\citenamefont {Duan}\ \emph {et~al.}()\citenamefont {Duan} \emph {et~al.}}]{duan2019}%
  \BibitemOpen
  \bibfield  {author} {\bibinfo {author} {\bibfnamefont {Z.}~\bibnamefont {Duan}} \emph {et~al.},\ }\bibfield  {title} {{\selectlanguage {english}\bibinfo {title} {{Concepts of Longitudinally Polarized Electron and Positron Colliding Beams in the Circular Electron Positron Collider}}},\ }in\ \href {https://doi.org/10.18429/JACoW-IPAC2019-MOPMP012} {{\selectlanguage {english}\emph {\bibinfo {booktitle} {Proc. IPAC'19}}}}\ (\bibinfo  {publisher} {JACoW Publishing, Geneva, Switzerland})\ pp.\ \bibinfo {pages} {445--448}\BibitemShut {NoStop}%
\bibitem [{\citenamefont {Steffens}\ \emph {et~al.}(1993)\citenamefont {Steffens}, \citenamefont {Andresen}, \citenamefont {Blume-Werry}, \citenamefont {Klein}, \citenamefont {Aulenbacher},\ and\ \citenamefont {Reichert}}]{Steffens1993}%
  \BibitemOpen
  \bibfield  {author} {\bibinfo {author} {\bibfnamefont {K.-H.}\ \bibnamefont {Steffens}}, \bibinfo {author} {\bibfnamefont {H.}~\bibnamefont {Andresen}}, \bibinfo {author} {\bibfnamefont {J.}~\bibnamefont {Blume-Werry}}, \bibinfo {author} {\bibfnamefont {F.}~\bibnamefont {Klein}}, \bibinfo {author} {\bibfnamefont {K.}~\bibnamefont {Aulenbacher}},\ and\ \bibinfo {author} {\bibfnamefont {E.}~\bibnamefont {Reichert}},\ }\bibfield  {title} {\bibinfo {title} {A spin rotator for producing a longitudinally polarized electron beam with mami},\ }\href {https://doi.org/https://doi.org/10.1016/0168-9002(93)90383-S} {\bibfield  {journal} {\bibinfo  {journal} {Nuclear Instruments and Methods in Physics Research Section A: Accelerators, Spectrometers, Detectors and Associated Equipment}\ }\textbf {\bibinfo {volume} {325}},\ \bibinfo {pages} {378} (\bibinfo {year} {1993})}\BibitemShut {NoStop}%
\bibitem [{\citenamefont {Buon}\ and\ \citenamefont {Steffen}(1986)}]{Buon1986}%
  \BibitemOpen
  \bibfield  {author} {\bibinfo {author} {\bibfnamefont {J.}~\bibnamefont {Buon}}\ and\ \bibinfo {author} {\bibfnamefont {K.}~\bibnamefont {Steffen}},\ }\bibfield  {title} {\bibinfo {title} {Hera variable-energy “mini” spin rotator and head-on ep collision scheme with choice of electron helicity},\ }\href {https://doi.org/https://doi.org/10.1016/0168-9002(86)91257-X} {\bibfield  {journal} {\bibinfo  {journal} {Nuclear Instruments and Methods in Physics Research Section A: Accelerators, Spectrometers, Detectors and Associated Equipment}\ }\textbf {\bibinfo {volume} {245}},\ \bibinfo {pages} {248} (\bibinfo {year} {1986})}\BibitemShut {NoStop}%
\bibitem [{\citenamefont {Li}\ \emph {et~al.}(2025)\citenamefont {Li}, \citenamefont {Wang}, \citenamefont {Sun}, \citenamefont {Wan}, \citenamefont {Salamin}, \citenamefont {Ababekri}, \citenamefont {Zhao}, \citenamefont {Xue}, \citenamefont {Tian}, \citenamefont {Wei},\ and\ \citenamefont {Li}}]{Li2025}%
  \BibitemOpen
  \bibfield  {author} {\bibinfo {author} {\bibfnamefont {Z.-P.}\ \bibnamefont {Li}}, \bibinfo {author} {\bibfnamefont {Y.}~\bibnamefont {Wang}}, \bibinfo {author} {\bibfnamefont {T.}~\bibnamefont {Sun}}, \bibinfo {author} {\bibfnamefont {F.}~\bibnamefont {Wan}}, \bibinfo {author} {\bibfnamefont {Y.~I.}\ \bibnamefont {Salamin}}, \bibinfo {author} {\bibfnamefont {M.}~\bibnamefont {Ababekri}}, \bibinfo {author} {\bibfnamefont {Q.}~\bibnamefont {Zhao}}, \bibinfo {author} {\bibfnamefont {K.}~\bibnamefont {Xue}}, \bibinfo {author} {\bibfnamefont {Y.}~\bibnamefont {Tian}}, \bibinfo {author} {\bibfnamefont {W.-Q.}\ \bibnamefont {Wei}},\ and\ \bibinfo {author} {\bibfnamefont {J.-X.}\ \bibnamefont {Li}},\ }\bibfield  {title} {\bibinfo {title} {Ultrafast spin rotation of relativistic lepton beams via terahertz wave in a dielectric-lined waveguide},\ }\href {https://doi.org/10.1103/PhysRevLett.134.075001} {\bibfield  {journal} {\bibinfo  {journal} {Phys. Rev. Lett.}\ }\textbf {\bibinfo {volume} {134}},\ \bibinfo {pages}
  {075001} (\bibinfo {year} {2025})}\BibitemShut {NoStop}%
\bibitem [{\citenamefont {Li}\ \emph {et~al.}(2020{\natexlab{a}})\citenamefont {Li}, \citenamefont {Chen}, \citenamefont {Wang},\ and\ \citenamefont {Hu}}]{li2020production}%
  \BibitemOpen
  \bibfield  {author} {\bibinfo {author} {\bibfnamefont {Y.-F.}\ \bibnamefont {Li}}, \bibinfo {author} {\bibfnamefont {Y.-Y.}\ \bibnamefont {Chen}}, \bibinfo {author} {\bibfnamefont {W.-M.}\ \bibnamefont {Wang}},\ and\ \bibinfo {author} {\bibfnamefont {H.-S.}\ \bibnamefont {Hu}},\ }\bibfield  {title} {\bibinfo {title} {Production of highly polarized positron beams via helicity transfer from polarized electrons in a strong laser field},\ }\href@noop {} {\bibfield  {journal} {\bibinfo  {journal} {Physical Review Letters}\ }\textbf {\bibinfo {volume} {125}},\ \bibinfo {pages} {044802} (\bibinfo {year} {2020}{\natexlab{a}})}\BibitemShut {NoStop}%
\bibitem [{\citenamefont {Wen}\ \emph {et~al.}(2019)\citenamefont {Wen}, \citenamefont {Tamburini},\ and\ \citenamefont {Keitel}}]{wen2019}%
  \BibitemOpen
  \bibfield  {author} {\bibinfo {author} {\bibfnamefont {M.}~\bibnamefont {Wen}}, \bibinfo {author} {\bibfnamefont {M.}~\bibnamefont {Tamburini}},\ and\ \bibinfo {author} {\bibfnamefont {C.~H.}\ \bibnamefont {Keitel}},\ }\bibfield  {title} {\bibinfo {title} {Polarized laser-wakefield-accelerated kiloampere electron beams},\ }\href {https://doi.org/10.1103/PhysRevLett.122.214801} {\bibfield  {journal} {\bibinfo  {journal} {Phys. Rev. Lett.}\ }\textbf {\bibinfo {volume} {122}},\ \bibinfo {pages} {214801} (\bibinfo {year} {2019})}\BibitemShut {NoStop}%
\bibitem [{\citenamefont {Li}\ \emph {et~al.}(2019)\citenamefont {Li}, \citenamefont {Shaisultanov}, \citenamefont {Hatsagortsyan}, \citenamefont {Wan}, \citenamefont {Keitel},\ and\ \citenamefont {Li}}]{li2019ultrarelativistic}%
  \BibitemOpen
  \bibfield  {author} {\bibinfo {author} {\bibfnamefont {Y.-F.}\ \bibnamefont {Li}}, \bibinfo {author} {\bibfnamefont {R.}~\bibnamefont {Shaisultanov}}, \bibinfo {author} {\bibfnamefont {K.~Z.}\ \bibnamefont {Hatsagortsyan}}, \bibinfo {author} {\bibfnamefont {F.}~\bibnamefont {Wan}}, \bibinfo {author} {\bibfnamefont {C.~H.}\ \bibnamefont {Keitel}},\ and\ \bibinfo {author} {\bibfnamefont {J.-X.}\ \bibnamefont {Li}},\ }\bibfield  {title} {\bibinfo {title} {Ultrarelativistic electron-beam polarization in single-shot interaction with an ultraintense laser pulse},\ }\href@noop {} {\bibfield  {journal} {\bibinfo  {journal} {Physical review letters}\ }\textbf {\bibinfo {volume} {122}},\ \bibinfo {pages} {154801} (\bibinfo {year} {2019})}\BibitemShut {NoStop}%
\bibitem [{\citenamefont {Liu}\ \emph {et~al.}(2022)\citenamefont {Liu}, \citenamefont {Xue}, \citenamefont {Wan}, \citenamefont {Chen}, \citenamefont {Li}, \citenamefont {Liu}, \citenamefont {Weng}, \citenamefont {Sheng},\ and\ \citenamefont {Zhang}}]{Liu2022}%
  \BibitemOpen
  \bibfield  {author} {\bibinfo {author} {\bibfnamefont {W.-Y.}\ \bibnamefont {Liu}}, \bibinfo {author} {\bibfnamefont {K.}~\bibnamefont {Xue}}, \bibinfo {author} {\bibfnamefont {F.}~\bibnamefont {Wan}}, \bibinfo {author} {\bibfnamefont {M.}~\bibnamefont {Chen}}, \bibinfo {author} {\bibfnamefont {J.-X.}\ \bibnamefont {Li}}, \bibinfo {author} {\bibfnamefont {F.}~\bibnamefont {Liu}}, \bibinfo {author} {\bibfnamefont {S.-M.}\ \bibnamefont {Weng}}, \bibinfo {author} {\bibfnamefont {Z.-M.}\ \bibnamefont {Sheng}},\ and\ \bibinfo {author} {\bibfnamefont {J.}~\bibnamefont {Zhang}},\ }\bibfield  {title} {\bibinfo {title} {Trapping and acceleration of spin-polarized positrons from $\ensuremath{\gamma}$ photon splitting in wakefields},\ }\href {https://doi.org/10.1103/PhysRevResearch.4.L022028} {\bibfield  {journal} {\bibinfo  {journal} {Phys. Rev. Res.}\ }\textbf {\bibinfo {volume} {4}},\ \bibinfo {pages} {L022028} (\bibinfo {year} {2022})}\BibitemShut {NoStop}%
\bibitem [{\citenamefont {Lu}\ \emph {et~al.}(2006)\citenamefont {Lu}, \citenamefont {Huang}, \citenamefont {Zhou}, \citenamefont {Mori},\ and\ \citenamefont {Katsouleas}}]{Lu2006}%
  \BibitemOpen
  \bibfield  {author} {\bibinfo {author} {\bibfnamefont {W.}~\bibnamefont {Lu}}, \bibinfo {author} {\bibfnamefont {C.}~\bibnamefont {Huang}}, \bibinfo {author} {\bibfnamefont {M.}~\bibnamefont {Zhou}}, \bibinfo {author} {\bibfnamefont {W.~B.}\ \bibnamefont {Mori}},\ and\ \bibinfo {author} {\bibfnamefont {T.}~\bibnamefont {Katsouleas}},\ }\bibfield  {title} {\bibinfo {title} {Nonlinear theory for relativistic plasma wakefields in the blowout regime},\ }\href {https://doi.org/10.1103/PhysRevLett.96.165002} {\bibfield  {journal} {\bibinfo  {journal} {Phys. Rev. Lett.}\ }\textbf {\bibinfo {volume} {96}},\ \bibinfo {pages} {165002} (\bibinfo {year} {2006})}\BibitemShut {NoStop}%
\bibitem [{\citenamefont {Esarey}\ \emph {et~al.}(2009)\citenamefont {Esarey}, \citenamefont {Schroeder},\ and\ \citenamefont {Leemans}}]{RevModPhys2009}%
  \BibitemOpen
  \bibfield  {author} {\bibinfo {author} {\bibfnamefont {E.}~\bibnamefont {Esarey}}, \bibinfo {author} {\bibfnamefont {C.~B.}\ \bibnamefont {Schroeder}},\ and\ \bibinfo {author} {\bibfnamefont {W.~P.}\ \bibnamefont {Leemans}},\ }\bibfield  {title} {\bibinfo {title} {Physics of laser-driven plasma-based electron accelerators},\ }\href {https://doi.org/10.1103/RevModPhys.81.1229} {\bibfield  {journal} {\bibinfo  {journal} {Rev. Mod. Phys.}\ }\textbf {\bibinfo {volume} {81}},\ \bibinfo {pages} {1229} (\bibinfo {year} {2009})}\BibitemShut {NoStop}%
\bibitem [{\citenamefont {Pukhov}\ \emph {et~al.}(1999)\citenamefont {Pukhov}, \citenamefont {Sheng},\ and\ \citenamefont {Meyer-ter Vehn}}]{Pukhov1999}%
  \BibitemOpen
  \bibfield  {author} {\bibinfo {author} {\bibfnamefont {A.}~\bibnamefont {Pukhov}}, \bibinfo {author} {\bibfnamefont {Z.-M.}\ \bibnamefont {Sheng}},\ and\ \bibinfo {author} {\bibfnamefont {J.}~\bibnamefont {Meyer-ter Vehn}},\ }\bibfield  {title} {\bibinfo {title} {Particle acceleration in relativistic laser channels},\ }\href {https://doi.org/10.1063/1.873242} {\bibfield  {journal} {\bibinfo  {journal} {Physics of Plasmas}\ }\textbf {\bibinfo {volume} {6}},\ \bibinfo {pages} {2847} (\bibinfo {year} {1999})}\BibitemShut {NoStop}%
\bibitem [{\citenamefont {Gonoskov}\ \emph {et~al.}(2022)\citenamefont {Gonoskov}, \citenamefont {Blackburn}, \citenamefont {Marklund},\ and\ \citenamefont {Bulanov}}]{Gonoskov2022}%
  \BibitemOpen
  \bibfield  {author} {\bibinfo {author} {\bibfnamefont {A.}~\bibnamefont {Gonoskov}}, \bibinfo {author} {\bibfnamefont {T.~G.}\ \bibnamefont {Blackburn}}, \bibinfo {author} {\bibfnamefont {M.}~\bibnamefont {Marklund}},\ and\ \bibinfo {author} {\bibfnamefont {S.~S.}\ \bibnamefont {Bulanov}},\ }\bibfield  {title} {\bibinfo {title} {Charged particle motion and radiation in strong electromagnetic fields},\ }\href {https://doi.org/10.1103/RevModPhys.94.045001} {\bibfield  {journal} {\bibinfo  {journal} {Rev. Mod. Phys.}\ }\textbf {\bibinfo {volume} {94}},\ \bibinfo {pages} {045001} (\bibinfo {year} {2022})}\BibitemShut {NoStop}%
\bibitem [{\citenamefont {Martinez}\ \emph {et~al.}(2023)\citenamefont {Martinez}, \citenamefont {Barbosa},\ and\ \citenamefont {Vranic}}]{Martinez2023}%
  \BibitemOpen
  \bibfield  {author} {\bibinfo {author} {\bibfnamefont {B.}~\bibnamefont {Martinez}}, \bibinfo {author} {\bibfnamefont {B.}~\bibnamefont {Barbosa}},\ and\ \bibinfo {author} {\bibfnamefont {M.}~\bibnamefont {Vranic}},\ }\bibfield  {title} {\bibinfo {title} {Creation and direct laser acceleration of positrons in a single stage},\ }\href {https://doi.org/10.1103/PhysRevAccelBeams.26.011301} {\bibfield  {journal} {\bibinfo  {journal} {Phys. Rev. Accel. Beams}\ }\textbf {\bibinfo {volume} {26}},\ \bibinfo {pages} {011301} (\bibinfo {year} {2023})}\BibitemShut {NoStop}%
\bibitem [{Wu2(2025)}]{Wu2025}%
  \BibitemOpen
  \bibfield  {title} {\bibinfo {title} {Achieving high polarization of photons emitted by unpolarized electrons in ultrastrong laser fields},\ }\bibfield  {journal} {\bibinfo  {journal} {Communications Physics}\ }\textbf {\bibinfo {volume} {8}},\ \href {https://doi.org/10.1038/s42005-025-02077-2} {10.1038/s42005-025-02077-2} (\bibinfo {year} {2025})\BibitemShut {NoStop}%
\bibitem [{\citenamefont {Li}\ \emph {et~al.}(2020{\natexlab{b}})\citenamefont {Li}, \citenamefont {Shaisultanov}, \citenamefont {Chen}, \citenamefont {Wan}, \citenamefont {Hatsagortsyan}, \citenamefont {Keitel},\ and\ \citenamefont {Li}}]{li2020polarized}%
  \BibitemOpen
  \bibfield  {author} {\bibinfo {author} {\bibfnamefont {Y.-F.}\ \bibnamefont {Li}}, \bibinfo {author} {\bibfnamefont {R.}~\bibnamefont {Shaisultanov}}, \bibinfo {author} {\bibfnamefont {Y.-Y.}\ \bibnamefont {Chen}}, \bibinfo {author} {\bibfnamefont {F.}~\bibnamefont {Wan}}, \bibinfo {author} {\bibfnamefont {K.~Z.}\ \bibnamefont {Hatsagortsyan}}, \bibinfo {author} {\bibfnamefont {C.~H.}\ \bibnamefont {Keitel}},\ and\ \bibinfo {author} {\bibfnamefont {J.-X.}\ \bibnamefont {Li}},\ }\bibfield  {title} {\bibinfo {title} {Polarized ultrashort brilliant multi-gev $\gamma$ rays via single-shot laser-electron interaction},\ }\href@noop {} {\bibfield  {journal} {\bibinfo  {journal} {Physical review letters}\ }\textbf {\bibinfo {volume} {124}},\ \bibinfo {pages} {014801} (\bibinfo {year} {2020}{\natexlab{b}})}\BibitemShut {NoStop}%
\bibitem [{\citenamefont {Baier}\ \emph {et~al.}(1998)\citenamefont {Baier}, \citenamefont {Katkov},\ and\ \citenamefont {Strakhovenko}}]{baier1998electromagnetic}%
  \BibitemOpen
  \bibfield  {author} {\bibinfo {author} {\bibfnamefont {V.~N.}\ \bibnamefont {Baier}}, \bibinfo {author} {\bibfnamefont {V.~M.}\ \bibnamefont {Katkov}},\ and\ \bibinfo {author} {\bibfnamefont {V.~M.}\ \bibnamefont {Strakhovenko}},\ }\href@noop {} {\emph {\bibinfo {title} {Electromagnetic processes at high energies in oriented single crystals}}}\ (\bibinfo  {publisher} {World Scientific},\ \bibinfo {year} {1998})\BibitemShut {NoStop}%
\bibitem [{\citenamefont {Ritus}(1985)}]{ritus1985quantum}%
  \BibitemOpen
  \bibfield  {author} {\bibinfo {author} {\bibfnamefont {V.}~\bibnamefont {Ritus}},\ }\bibfield  {title} {\bibinfo {title} {Quantum effects of the interaction of elementary particles with an intense electromagnetic field},\ }\href@noop {} {\bibfield  {journal} {\bibinfo  {journal} {J. Sov. Laser Res.;(United States)}\ }\textbf {\bibinfo {volume} {6}} (\bibinfo {year} {1985})}\BibitemShut {NoStop}%
\bibitem [{\citenamefont {Di~Piazza}\ \emph {et~al.}(2018)\citenamefont {Di~Piazza}, \citenamefont {Tamburini}, \citenamefont {Meuren},\ and\ \citenamefont {Keitel}}]{di2018implementing}%
  \BibitemOpen
  \bibfield  {author} {\bibinfo {author} {\bibfnamefont {A.}~\bibnamefont {Di~Piazza}}, \bibinfo {author} {\bibfnamefont {M.}~\bibnamefont {Tamburini}}, \bibinfo {author} {\bibfnamefont {S.}~\bibnamefont {Meuren}},\ and\ \bibinfo {author} {\bibfnamefont {C.}~\bibnamefont {Keitel}},\ }\bibfield  {title} {\bibinfo {title} {Implementing nonlinear compton scattering beyond the local-constant-field approximation},\ }\href@noop {} {\bibfield  {journal} {\bibinfo  {journal} {Physical Review A}\ }\textbf {\bibinfo {volume} {98}},\ \bibinfo {pages} {012134} (\bibinfo {year} {2018})}\BibitemShut {NoStop}%
\bibitem [{\citenamefont {Di~Piazza}\ \emph {et~al.}(2019)\citenamefont {Di~Piazza}, \citenamefont {Tamburini}, \citenamefont {Meuren},\ and\ \citenamefont {Keitel}}]{di2019improved}%
  \BibitemOpen
  \bibfield  {author} {\bibinfo {author} {\bibfnamefont {A.}~\bibnamefont {Di~Piazza}}, \bibinfo {author} {\bibfnamefont {M.}~\bibnamefont {Tamburini}}, \bibinfo {author} {\bibfnamefont {S.}~\bibnamefont {Meuren}},\ and\ \bibinfo {author} {\bibfnamefont {C.~H.}\ \bibnamefont {Keitel}},\ }\bibfield  {title} {\bibinfo {title} {Improved local-constant-field approximation for strong-field qed codes},\ }\href@noop {} {\bibfield  {journal} {\bibinfo  {journal} {Physical Review A}\ }\textbf {\bibinfo {volume} {99}},\ \bibinfo {pages} {022125} (\bibinfo {year} {2019})}\BibitemShut {NoStop}%
\bibitem [{\citenamefont {Gonsalves}\ \emph {et~al.}(2019)\citenamefont {Gonsalves}, \citenamefont {Nakamura}, \citenamefont {Daniels}, \citenamefont {Benedetti}, \citenamefont {Pieronek}, \citenamefont {De~Raadt}, \citenamefont {Steinke}, \citenamefont {Bin}, \citenamefont {Bulanov}, \citenamefont {Van~Tilborg} \emph {et~al.}}]{gonsalves2019petawatt}%
  \BibitemOpen
  \bibfield  {author} {\bibinfo {author} {\bibfnamefont {A.}~\bibnamefont {Gonsalves}}, \bibinfo {author} {\bibfnamefont {K.}~\bibnamefont {Nakamura}}, \bibinfo {author} {\bibfnamefont {J.}~\bibnamefont {Daniels}}, \bibinfo {author} {\bibfnamefont {C.}~\bibnamefont {Benedetti}}, \bibinfo {author} {\bibfnamefont {C.}~\bibnamefont {Pieronek}}, \bibinfo {author} {\bibfnamefont {T.}~\bibnamefont {De~Raadt}}, \bibinfo {author} {\bibfnamefont {S.}~\bibnamefont {Steinke}}, \bibinfo {author} {\bibfnamefont {J.}~\bibnamefont {Bin}}, \bibinfo {author} {\bibfnamefont {S.}~\bibnamefont {Bulanov}}, \bibinfo {author} {\bibfnamefont {J.}~\bibnamefont {Van~Tilborg}}, \emph {et~al.},\ }\bibfield  {title} {\bibinfo {title} {Petawatt laser guiding and electron beam acceleration to 8 gev in a laser-heated capillary discharge waveguide},\ }\href@noop {} {\bibfield  {journal} {\bibinfo  {journal} {Physical review letters}\ }\textbf {\bibinfo {volume} {122}},\ \bibinfo {pages} {084801} (\bibinfo {year} {2019})}\BibitemShut
  {NoStop}%
\bibitem [{\citenamefont {Leemans}\ \emph {et~al.}(2014)\citenamefont {Leemans}, \citenamefont {Gonsalves}, \citenamefont {Mao}, \citenamefont {Nakamura}, \citenamefont {Benedetti}, \citenamefont {Schroeder}, \citenamefont {T{\'o}th}, \citenamefont {Daniels}, \citenamefont {Mittelberger}, \citenamefont {Bulanov} \emph {et~al.}}]{leemans2014multi}%
  \BibitemOpen
  \bibfield  {author} {\bibinfo {author} {\bibfnamefont {W.}~\bibnamefont {Leemans}}, \bibinfo {author} {\bibfnamefont {A.}~\bibnamefont {Gonsalves}}, \bibinfo {author} {\bibfnamefont {H.-S.}\ \bibnamefont {Mao}}, \bibinfo {author} {\bibfnamefont {K.}~\bibnamefont {Nakamura}}, \bibinfo {author} {\bibfnamefont {C.}~\bibnamefont {Benedetti}}, \bibinfo {author} {\bibfnamefont {C.}~\bibnamefont {Schroeder}}, \bibinfo {author} {\bibfnamefont {C.}~\bibnamefont {T{\'o}th}}, \bibinfo {author} {\bibfnamefont {J.}~\bibnamefont {Daniels}}, \bibinfo {author} {\bibfnamefont {D.}~\bibnamefont {Mittelberger}}, \bibinfo {author} {\bibfnamefont {S.}~\bibnamefont {Bulanov}}, \emph {et~al.},\ }\bibfield  {title} {\bibinfo {title} {Multi-gev electron beams from capillary-discharge-guided subpetawatt laser pulses in the self-trapping regime},\ }\href@noop {} {\bibfield  {journal} {\bibinfo  {journal} {Physical review letters}\ }\textbf {\bibinfo {volume} {113}},\ \bibinfo {pages} {245002} (\bibinfo {year} {2014})}\BibitemShut
  {NoStop}%
\bibitem [{SM()}]{SM}%
  \BibitemOpen
  \href@noop {} {}\bibinfo {note} {See Supplemental Information for details on the simulated results for other laser and electron parameters, and theortical analysis of positron polarization.}\BibitemShut {Stop}%
\bibitem [{\citenamefont {Yoon}\ \emph {et~al.}(2021)\citenamefont {Yoon}, \citenamefont {Kim}, \citenamefont {Choi}, \citenamefont {Sung}, \citenamefont {Lee}, \citenamefont {Lee},\ and\ \citenamefont {Nam}}]{yoon2021realization}%
  \BibitemOpen
  \bibfield  {author} {\bibinfo {author} {\bibfnamefont {J.~W.}\ \bibnamefont {Yoon}}, \bibinfo {author} {\bibfnamefont {Y.~G.}\ \bibnamefont {Kim}}, \bibinfo {author} {\bibfnamefont {I.~W.}\ \bibnamefont {Choi}}, \bibinfo {author} {\bibfnamefont {J.~H.}\ \bibnamefont {Sung}}, \bibinfo {author} {\bibfnamefont {H.~W.}\ \bibnamefont {Lee}}, \bibinfo {author} {\bibfnamefont {S.~K.}\ \bibnamefont {Lee}},\ and\ \bibinfo {author} {\bibfnamefont {C.~H.}\ \bibnamefont {Nam}},\ }\bibfield  {title} {\bibinfo {title} {Realization of laser intensity over 1023 w/cm2},\ }\href@noop {} {\bibfield  {journal} {\bibinfo  {journal} {Optica}\ }\textbf {\bibinfo {volume} {8}},\ \bibinfo {pages} {630} (\bibinfo {year} {2021})}\BibitemShut {NoStop}%
\bibitem [{\citenamefont {Danson}\ \emph {et~al.}(2019)\citenamefont {Danson}, \citenamefont {Haefner}, \citenamefont {Bromage}, \citenamefont {Butcher}, \citenamefont {Chanteloup}, \citenamefont {Chowdhury}, \citenamefont {Galvanauskas}, \citenamefont {Gizzi}, \citenamefont {Hein}, \citenamefont {Hillier} \emph {et~al.}}]{danson2019petawatt}%
  \BibitemOpen
  \bibfield  {author} {\bibinfo {author} {\bibfnamefont {C.~N.}\ \bibnamefont {Danson}}, \bibinfo {author} {\bibfnamefont {C.}~\bibnamefont {Haefner}}, \bibinfo {author} {\bibfnamefont {J.}~\bibnamefont {Bromage}}, \bibinfo {author} {\bibfnamefont {T.}~\bibnamefont {Butcher}}, \bibinfo {author} {\bibfnamefont {J.-C.~F.}\ \bibnamefont {Chanteloup}}, \bibinfo {author} {\bibfnamefont {E.~A.}\ \bibnamefont {Chowdhury}}, \bibinfo {author} {\bibfnamefont {A.}~\bibnamefont {Galvanauskas}}, \bibinfo {author} {\bibfnamefont {L.~A.}\ \bibnamefont {Gizzi}}, \bibinfo {author} {\bibfnamefont {J.}~\bibnamefont {Hein}}, \bibinfo {author} {\bibfnamefont {D.~I.}\ \bibnamefont {Hillier}}, \emph {et~al.},\ }\bibfield  {title} {\bibinfo {title} {Petawatt and exawatt class lasers worldwide},\ }\href@noop {} {\bibfield  {journal} {\bibinfo  {journal} {High Power Laser Science and Engineering}\ }\textbf {\bibinfo {volume} {7}},\ \bibinfo {pages} {e54} (\bibinfo {year} {2019})}\BibitemShut {NoStop}%
\bibitem [{\citenamefont {Li}\ \emph {et~al.}(2023)\citenamefont {Li}, \citenamefont {Li}, \citenamefont {Chen}, \citenamefont {Weng}, \citenamefont {Tan}, \citenamefont {Ma}, \citenamefont {Sheng},\ and\ \citenamefont {Hu}}]{li2023highly}%
  \BibitemOpen
  \bibfield  {author} {\bibinfo {author} {\bibfnamefont {B.-J.}\ \bibnamefont {Li}}, \bibinfo {author} {\bibfnamefont {Y.-F.}\ \bibnamefont {Li}}, \bibinfo {author} {\bibfnamefont {Y.-Y.}\ \bibnamefont {Chen}}, \bibinfo {author} {\bibfnamefont {X.-F.}\ \bibnamefont {Weng}}, \bibinfo {author} {\bibfnamefont {X.-J.}\ \bibnamefont {Tan}}, \bibinfo {author} {\bibfnamefont {X.-J.}\ \bibnamefont {Ma}}, \bibinfo {author} {\bibfnamefont {L.}~\bibnamefont {Sheng}},\ and\ \bibinfo {author} {\bibfnamefont {H.-S.}\ \bibnamefont {Hu}},\ }\bibfield  {title} {\bibinfo {title} {Highly polarized positrons generated via few-pw lasers},\ }\href@noop {} {\bibfield  {journal} {\bibinfo  {journal} {Physics of Plasmas}\ }\textbf {\bibinfo {volume} {30}} (\bibinfo {year} {2023})}\BibitemShut {NoStop}%
\bibitem [{\citenamefont {Jackson}(1998)}]{jackson1998classical}%
  \BibitemOpen
  \bibfield  {author} {\bibinfo {author} {\bibfnamefont {J.~D.}\ \bibnamefont {Jackson}},\ }\href@noop {} {\emph {\bibinfo {title} {Classical electrodynamics}}}\ (\bibinfo  {publisher} {John Wiley \& Sons},\ \bibinfo {year} {1998})\BibitemShut {NoStop}%
\bibitem [{\citenamefont {Ma}\ \emph {et~al.}(2018{\natexlab{a}})\citenamefont {Ma}, \citenamefont {Zhao}, \citenamefont {Li}, \citenamefont {Li}, \citenamefont {Chen}, \citenamefont {Liu}, \citenamefont {Dann}, \citenamefont {Ma}, \citenamefont {Yang}, \citenamefont {Ge}, \citenamefont {Sheng},\ and\ \citenamefont {Zhang}}]{Ma2018}%
  \BibitemOpen
  \bibfield  {author} {\bibinfo {author} {\bibfnamefont {Y.}~\bibnamefont {Ma}}, \bibinfo {author} {\bibfnamefont {J.}~\bibnamefont {Zhao}}, \bibinfo {author} {\bibfnamefont {Y.}~\bibnamefont {Li}}, \bibinfo {author} {\bibfnamefont {D.}~\bibnamefont {Li}}, \bibinfo {author} {\bibfnamefont {L.}~\bibnamefont {Chen}}, \bibinfo {author} {\bibfnamefont {J.}~\bibnamefont {Liu}}, \bibinfo {author} {\bibfnamefont {S.~J.~D.}\ \bibnamefont {Dann}}, \bibinfo {author} {\bibfnamefont {Y.}~\bibnamefont {Ma}}, \bibinfo {author} {\bibfnamefont {X.}~\bibnamefont {Yang}}, \bibinfo {author} {\bibfnamefont {Z.}~\bibnamefont {Ge}}, \bibinfo {author} {\bibfnamefont {Z.}~\bibnamefont {Sheng}},\ and\ \bibinfo {author} {\bibfnamefont {J.}~\bibnamefont {Zhang}},\ }\bibfield  {title} {\bibinfo {title} {Ultrahigh-charge electron beams from laser-irradiated solid surface},\ }\href {https://doi.org/10.1073/pnas.1800668115} {\bibfield  {journal} {\bibinfo  {journal} {Proceedings of the National Academy of Sciences}\ }\textbf {\bibinfo
  {volume} {115}},\ \bibinfo {pages} {6980} (\bibinfo {year} {2018}{\natexlab{a}})},\ \Eprint {https://arxiv.org/abs/https://www.pnas.org/doi/pdf/10.1073/pnas.1800668115} {https://www.pnas.org/doi/pdf/10.1073/pnas.1800668115} \BibitemShut {NoStop}%
\bibitem [{sec()}]{section}%
  \BibitemOpen
  \href@noop {} {}\bibinfo {howpublished} {\url{https://www.osti.gov/biblio/954154/}}\BibitemShut {NoStop}%
\bibitem [{\citenamefont {Van~House}\ and\ \citenamefont {Zitzewitz}(1984)}]{van1984probing}%
  \BibitemOpen
  \bibfield  {author} {\bibinfo {author} {\bibfnamefont {J.}~\bibnamefont {Van~House}}\ and\ \bibinfo {author} {\bibfnamefont {P.}~\bibnamefont {Zitzewitz}},\ }\bibfield  {title} {\bibinfo {title} {Probing the positron moderation process using high-intensity, highly polarized slow-positron beams},\ }\href@noop {} {\bibfield  {journal} {\bibinfo  {journal} {Physical Review A}\ }\textbf {\bibinfo {volume} {29}},\ \bibinfo {pages} {96} (\bibinfo {year} {1984})}\BibitemShut {NoStop}%
\bibitem [{\citenamefont {Subashiev}\ \emph {et~al.}(1998)\citenamefont {Subashiev}, \citenamefont {Yashin}, \citenamefont {Clendenin},\ and\ \citenamefont {Mamaev}}]{subashiev1998spin}%
  \BibitemOpen
  \bibfield  {author} {\bibinfo {author} {\bibfnamefont {A.}~\bibnamefont {Subashiev}}, \bibinfo {author} {\bibfnamefont {Y.~P.}\ \bibnamefont {Yashin}}, \bibinfo {author} {\bibfnamefont {J.}~\bibnamefont {Clendenin}},\ and\ \bibinfo {author} {\bibfnamefont {Y.~A.}\ \bibnamefont {Mamaev}},\ }\bibfield  {title} {\bibinfo {title} {Spin polarized electrons: Generation and applications},\ }\href@noop {} {\bibfield  {journal} {\bibinfo  {journal} {Phys. Low Dimens. Struct}\ }\textbf {\bibinfo {volume} {1}} (\bibinfo {year} {1998})}\BibitemShut {NoStop}%
\bibitem [{\citenamefont {Gidley}\ \emph {et~al.}(1982)\citenamefont {Gidley}, \citenamefont {K{\"o}ymen},\ and\ \citenamefont {Capehart}}]{gidley1982polarized}%
  \BibitemOpen
  \bibfield  {author} {\bibinfo {author} {\bibfnamefont {D.}~\bibnamefont {Gidley}}, \bibinfo {author} {\bibfnamefont {A.}~\bibnamefont {K{\"o}ymen}},\ and\ \bibinfo {author} {\bibfnamefont {T.~W.}\ \bibnamefont {Capehart}},\ }\bibfield  {title} {\bibinfo {title} {Polarized low-energy positrons: A new probe of surface magnetism},\ }\href@noop {} {\bibfield  {journal} {\bibinfo  {journal} {Physical Review Letters}\ }\textbf {\bibinfo {volume} {49}},\ \bibinfo {pages} {1779} (\bibinfo {year} {1982})}\BibitemShut {NoStop}%
\bibitem [{\citenamefont {McMaster}(1961)}]{mcmaster1961matrix}%
  \BibitemOpen
  \bibfield  {author} {\bibinfo {author} {\bibfnamefont {W.~H.}\ \bibnamefont {McMaster}},\ }\bibfield  {title} {\bibinfo {title} {Matrix representation of polarization},\ }\href@noop {} {\bibfield  {journal} {\bibinfo  {journal} {Reviews of modern physics}\ }\textbf {\bibinfo {volume} {33}},\ \bibinfo {pages} {8} (\bibinfo {year} {1961})}\BibitemShut {NoStop}%
\bibitem [{\citenamefont {Baskov}(2015)}]{baskov2015radiation}%
  \BibitemOpen
  \bibfield  {author} {\bibinfo {author} {\bibfnamefont {V.}~\bibnamefont {Baskov}},\ }\bibfield  {title} {\bibinfo {title} {Radiation length of the oriented crystal},\ }\href@noop {} {\bibfield  {journal} {\bibinfo  {journal} {Bulletin of the Lebedev Physics Institute}\ }\textbf {\bibinfo {volume} {42}},\ \bibinfo {pages} {144} (\bibinfo {year} {2015})}\BibitemShut {NoStop}%
\bibitem [{\citenamefont {Abbott}\ \emph {et~al.}(2016{\natexlab{b}})\citenamefont {Abbott}, \citenamefont {Adderley}, \citenamefont {Adeyemi}, \citenamefont {Aguilera}, \citenamefont {Ali}, \citenamefont {Areti}, \citenamefont {Baylac}, \citenamefont {Benesch}, \citenamefont {Bosson}, \citenamefont {Cade} \emph {et~al.}}]{abbott2016production}%
  \BibitemOpen
  \bibfield  {author} {\bibinfo {author} {\bibfnamefont {D.}~\bibnamefont {Abbott}}, \bibinfo {author} {\bibfnamefont {P.}~\bibnamefont {Adderley}}, \bibinfo {author} {\bibfnamefont {A.}~\bibnamefont {Adeyemi}}, \bibinfo {author} {\bibfnamefont {P.}~\bibnamefont {Aguilera}}, \bibinfo {author} {\bibfnamefont {M.}~\bibnamefont {Ali}}, \bibinfo {author} {\bibfnamefont {H.}~\bibnamefont {Areti}}, \bibinfo {author} {\bibfnamefont {M.}~\bibnamefont {Baylac}}, \bibinfo {author} {\bibfnamefont {J.}~\bibnamefont {Benesch}}, \bibinfo {author} {\bibfnamefont {G.}~\bibnamefont {Bosson}}, \bibinfo {author} {\bibfnamefont {B.}~\bibnamefont {Cade}}, \emph {et~al.},\ }\bibfield  {title} {\bibinfo {title} {Production of highly polarized positrons using polarized electrons at mev energies},\ }\href@noop {} {\bibfield  {journal} {\bibinfo  {journal} {Physical review letters}\ }\textbf {\bibinfo {volume} {116}},\ \bibinfo {pages} {214801} (\bibinfo {year} {2016}{\natexlab{b}})}\BibitemShut {NoStop}%
\bibitem [{\citenamefont {Kotkin}\ \emph {et~al.}(1998)\citenamefont {Kotkin}, \citenamefont {Perlt},\ and\ \citenamefont {Serbo}}]{kotkin1998polarization}%
  \BibitemOpen
  \bibfield  {author} {\bibinfo {author} {\bibfnamefont {G.}~\bibnamefont {Kotkin}}, \bibinfo {author} {\bibfnamefont {H.}~\bibnamefont {Perlt}},\ and\ \bibinfo {author} {\bibfnamefont {V.}~\bibnamefont {Serbo}},\ }\bibfield  {title} {\bibinfo {title} {Polarization of high-energy electrons traversing a laser beam},\ }\href@noop {} {\bibfield  {journal} {\bibinfo  {journal} {Nuclear Instruments and Methods in Physics Research Section A: Accelerators, Spectrometers, Detectors and Associated Equipment}\ }\textbf {\bibinfo {volume} {404}},\ \bibinfo {pages} {430} (\bibinfo {year} {1998})}\BibitemShut {NoStop}%
\bibitem [{\citenamefont {Clendenin}(2001)}]{clendenin2001recent}%
  \BibitemOpen
  \bibfield  {author} {\bibinfo {author} {\bibfnamefont {J.~E.}\ \bibnamefont {Clendenin}},\ }\bibfield  {title} {\bibinfo {title} {Recent advances in electron and positron sources},\ }in\ \href@noop {} {\emph {\bibinfo {booktitle} {AIP Conference Proceedings}}},\ Vol.\ \bibinfo {volume} {569}\ (\bibinfo {organization} {American Institute of Physics},\ \bibinfo {year} {2001})\ pp.\ \bibinfo {pages} {563--570}\BibitemShut {NoStop}%
\bibitem [{\citenamefont {A{\ss}mann}\ and\ \citenamefont {Zimmermann}(2001)}]{assmann2001polarization}%
  \BibitemOpen
  \bibfield  {author} {\bibinfo {author} {\bibfnamefont {R.~W.}\ \bibnamefont {A{\ss}mann}}\ and\ \bibinfo {author} {\bibfnamefont {F.}~\bibnamefont {Zimmermann}},\ }\href@noop {} {\emph {\bibinfo {title} {Polarization issues at CLIC}}},\ \bibinfo {type} {Tech. Rep.}\ (\bibinfo {year} {2001})\BibitemShut {NoStop}%
\bibitem [{\citenamefont {Kotkin}\ \emph {et~al.}(2003)\citenamefont {Kotkin}, \citenamefont {Serbo},\ and\ \citenamefont {Telnov}}]{kotkin2003electron}%
  \BibitemOpen
  \bibfield  {author} {\bibinfo {author} {\bibfnamefont {G.}~\bibnamefont {Kotkin}}, \bibinfo {author} {\bibfnamefont {V.}~\bibnamefont {Serbo}},\ and\ \bibinfo {author} {\bibfnamefont {V.~I.}\ \bibnamefont {Telnov}},\ }\bibfield  {title} {\bibinfo {title} {Electron (positron) beam polarization by compton scattering on circularly polarized laser photons},\ }\href@noop {} {\bibfield  {journal} {\bibinfo  {journal} {Physical Review Special Topics—Accelerators and Beams}\ }\textbf {\bibinfo {volume} {6}},\ \bibinfo {pages} {011001} (\bibinfo {year} {2003})}\BibitemShut {NoStop}%
\bibitem [{\citenamefont {Seipt}\ \emph {et~al.}(2018)\citenamefont {Seipt}, \citenamefont {Del~Sorbo}, \citenamefont {Ridgers},\ and\ \citenamefont {Thomas}}]{seipt2018theory}%
  \BibitemOpen
  \bibfield  {author} {\bibinfo {author} {\bibfnamefont {D.}~\bibnamefont {Seipt}}, \bibinfo {author} {\bibfnamefont {D.}~\bibnamefont {Del~Sorbo}}, \bibinfo {author} {\bibfnamefont {C.}~\bibnamefont {Ridgers}},\ and\ \bibinfo {author} {\bibfnamefont {A.}~\bibnamefont {Thomas}},\ }\bibfield  {title} {\bibinfo {title} {Theory of radiative electron polarization in strong laser fields},\ }\href@noop {} {\bibfield  {journal} {\bibinfo  {journal} {Physical Review A}\ }\textbf {\bibinfo {volume} {98}},\ \bibinfo {pages} {023417} (\bibinfo {year} {2018})}\BibitemShut {NoStop}%
\bibitem [{\citenamefont {Berestetskii}\ \emph {et~al.}(2012)\citenamefont {Berestetskii}, \citenamefont {Pitaevskii},\ and\ \citenamefont {Lifshitz}}]{berestetskii2012quantum}%
  \BibitemOpen
  \bibfield  {author} {\bibinfo {author} {\bibfnamefont {V.~B.}\ \bibnamefont {Berestetskii}}, \bibinfo {author} {\bibfnamefont {L.~P.}\ \bibnamefont {Pitaevskii}},\ and\ \bibinfo {author} {\bibfnamefont {E.~M.}\ \bibnamefont {Lifshitz}},\ }\href@noop {} {\emph {\bibinfo {title} {Quantum Electrodynamics: Volume 4}}},\ Vol.~\bibinfo {volume} {4}\ (\bibinfo  {publisher} {Elsevier},\ \bibinfo {year} {2012})\BibitemShut {NoStop}%
\bibitem [{\citenamefont {Di~Piazza}\ \emph {et~al.}(2012)\citenamefont {Di~Piazza}, \citenamefont {M{\"u}ller}, \citenamefont {Hatsagortsyan},\ and\ \citenamefont {Keitel}}]{di2012extremely}%
  \BibitemOpen
  \bibfield  {author} {\bibinfo {author} {\bibfnamefont {A.}~\bibnamefont {Di~Piazza}}, \bibinfo {author} {\bibfnamefont {C.}~\bibnamefont {M{\"u}ller}}, \bibinfo {author} {\bibfnamefont {K.}~\bibnamefont {Hatsagortsyan}},\ and\ \bibinfo {author} {\bibfnamefont {C.~H.}\ \bibnamefont {Keitel}},\ }\bibfield  {title} {\bibinfo {title} {Extremely high-intensity laser interactions with fundamental quantum systems},\ }\href@noop {} {\bibfield  {journal} {\bibinfo  {journal} {Reviews of Modern Physics}\ }\textbf {\bibinfo {volume} {84}},\ \bibinfo {pages} {1177} (\bibinfo {year} {2012})}\BibitemShut {NoStop}%
\bibitem [{\citenamefont {Ma}\ \emph {et~al.}(2018{\natexlab{b}})\citenamefont {Ma}, \citenamefont {Zhao}, \citenamefont {Li}, \citenamefont {Li}, \citenamefont {Chen}, \citenamefont {Liu}, \citenamefont {Dann}, \citenamefont {Ma}, \citenamefont {Yang}, \citenamefont {Ge} \emph {et~al.}}]{ma2018ultrahigh}%
  \BibitemOpen
  \bibfield  {author} {\bibinfo {author} {\bibfnamefont {Y.}~\bibnamefont {Ma}}, \bibinfo {author} {\bibfnamefont {J.}~\bibnamefont {Zhao}}, \bibinfo {author} {\bibfnamefont {Y.}~\bibnamefont {Li}}, \bibinfo {author} {\bibfnamefont {D.}~\bibnamefont {Li}}, \bibinfo {author} {\bibfnamefont {L.}~\bibnamefont {Chen}}, \bibinfo {author} {\bibfnamefont {J.}~\bibnamefont {Liu}}, \bibinfo {author} {\bibfnamefont {S.~J.}\ \bibnamefont {Dann}}, \bibinfo {author} {\bibfnamefont {Y.}~\bibnamefont {Ma}}, \bibinfo {author} {\bibfnamefont {X.}~\bibnamefont {Yang}}, \bibinfo {author} {\bibfnamefont {Z.}~\bibnamefont {Ge}}, \emph {et~al.},\ }\bibfield  {title} {\bibinfo {title} {Ultrahigh-charge electron beams from laser-irradiated solid surface},\ }\href@noop {} {\bibfield  {journal} {\bibinfo  {journal} {Proceedings of the National Academy of Sciences}\ }\textbf {\bibinfo {volume} {115}},\ \bibinfo {pages} {6980} (\bibinfo {year} {2018}{\natexlab{b}})}\BibitemShut {NoStop}%
\bibitem [{\citenamefont {Zhu}\ \emph {et~al.}(2016)\citenamefont {Zhu}, \citenamefont {Yu}, \citenamefont {Sheng}, \citenamefont {Yin}, \citenamefont {Turcu},\ and\ \citenamefont {Pukhov}}]{zhu2016dense}%
  \BibitemOpen
  \bibfield  {author} {\bibinfo {author} {\bibfnamefont {X.-L.}\ \bibnamefont {Zhu}}, \bibinfo {author} {\bibfnamefont {T.-P.}\ \bibnamefont {Yu}}, \bibinfo {author} {\bibfnamefont {Z.-M.}\ \bibnamefont {Sheng}}, \bibinfo {author} {\bibfnamefont {Y.}~\bibnamefont {Yin}}, \bibinfo {author} {\bibfnamefont {I.~C.~E.}\ \bibnamefont {Turcu}},\ and\ \bibinfo {author} {\bibfnamefont {A.}~\bibnamefont {Pukhov}},\ }\bibfield  {title} {\bibinfo {title} {Dense gev electron--positron pairs generated by lasers in near-critical-density plasmas},\ }\href@noop {} {\bibfield  {journal} {\bibinfo  {journal} {Nature communications}\ }\textbf {\bibinfo {volume} {7}},\ \bibinfo {pages} {13686} (\bibinfo {year} {2016})}\BibitemShut {NoStop}%
\bibitem [{\citenamefont {Weingartner}\ \emph {et~al.}(2012)\citenamefont {Weingartner}, \citenamefont {Raith}, \citenamefont {Popp}, \citenamefont {Chou}, \citenamefont {Wenz}, \citenamefont {Khrennikov}, \citenamefont {Heigoldt}, \citenamefont {Maier}, \citenamefont {Kajumba}, \citenamefont {Fuchs} \emph {et~al.}}]{weingartner2012ultralow}%
  \BibitemOpen
  \bibfield  {author} {\bibinfo {author} {\bibfnamefont {R.}~\bibnamefont {Weingartner}}, \bibinfo {author} {\bibfnamefont {S.}~\bibnamefont {Raith}}, \bibinfo {author} {\bibfnamefont {A.}~\bibnamefont {Popp}}, \bibinfo {author} {\bibfnamefont {S.}~\bibnamefont {Chou}}, \bibinfo {author} {\bibfnamefont {J.}~\bibnamefont {Wenz}}, \bibinfo {author} {\bibfnamefont {K.}~\bibnamefont {Khrennikov}}, \bibinfo {author} {\bibfnamefont {M.}~\bibnamefont {Heigoldt}}, \bibinfo {author} {\bibfnamefont {A.~R.}\ \bibnamefont {Maier}}, \bibinfo {author} {\bibfnamefont {N.}~\bibnamefont {Kajumba}}, \bibinfo {author} {\bibfnamefont {M.}~\bibnamefont {Fuchs}}, \emph {et~al.},\ }\bibfield  {title} {\bibinfo {title} {Ultralow emittance electron beams from a laser-wakefield accelerator},\ }\href@noop {} {\bibfield  {journal} {\bibinfo  {journal} {Physical Review Special Topics—Accelerators and Beams}\ }\textbf {\bibinfo {volume} {15}},\ \bibinfo {pages} {111302} (\bibinfo {year} {2012})}\BibitemShut {NoStop}%
\bibitem [{\citenamefont {Ivanov}\ \emph {et~al.}(2004)\citenamefont {Ivanov}, \citenamefont {Kotkin},\ and\ \citenamefont {Serbo}}]{ivanov2004complete}%
  \BibitemOpen
  \bibfield  {author} {\bibinfo {author} {\bibfnamefont {D.~Y.}\ \bibnamefont {Ivanov}}, \bibinfo {author} {\bibfnamefont {G.}~\bibnamefont {Kotkin}},\ and\ \bibinfo {author} {\bibfnamefont {V.}~\bibnamefont {Serbo}},\ }\bibfield  {title} {\bibinfo {title} {Complete description of polarization effects in emission of a photon by an electron in the field of a strong laser wave},\ }\href@noop {} {\bibfield  {journal} {\bibinfo  {journal} {The European Physical Journal C-Particles and Fields}\ }\textbf {\bibinfo {volume} {36}},\ \bibinfo {pages} {127} (\bibinfo {year} {2004})}\BibitemShut {NoStop}%
\bibitem [{\citenamefont {Karlovets}(2011)}]{karlovets2011radiative}%
  \BibitemOpen
  \bibfield  {author} {\bibinfo {author} {\bibfnamefont {D.~V.}\ \bibnamefont {Karlovets}},\ }\bibfield  {title} {\bibinfo {title} {Radiative polarization of electrons in a strong laser wave},\ }\href@noop {} {\bibfield  {journal} {\bibinfo  {journal} {Physical Review A—Atomic, Molecular, and Optical Physics}\ }\textbf {\bibinfo {volume} {84}},\ \bibinfo {pages} {062116} (\bibinfo {year} {2011})}\BibitemShut {NoStop}%
\bibitem [{\citenamefont {Song}\ \emph {et~al.}(2021)\citenamefont {Song}, \citenamefont {Wang},\ and\ \citenamefont {Li}}]{song2021generation}%
  \BibitemOpen
  \bibfield  {author} {\bibinfo {author} {\bibfnamefont {H.-H.}\ \bibnamefont {Song}}, \bibinfo {author} {\bibfnamefont {W.-M.}\ \bibnamefont {Wang}},\ and\ \bibinfo {author} {\bibfnamefont {Y.-T.}\ \bibnamefont {Li}},\ }\bibfield  {title} {\bibinfo {title} {Generation of polarized positron beams via collisions of ultrarelativistic electron beams},\ }\href@noop {} {\bibfield  {journal} {\bibinfo  {journal} {Physical Review Research}\ }\textbf {\bibinfo {volume} {3}},\ \bibinfo {pages} {033245} (\bibinfo {year} {2021})}\BibitemShut {NoStop}%
\bibitem [{\citenamefont {Chen}\ \emph {et~al.}(2022)\citenamefont {Chen}, \citenamefont {Hatsagortsyan}, \citenamefont {Keitel},\ and\ \citenamefont {Shaisultanov}}]{chen2022electron}%
  \BibitemOpen
  \bibfield  {author} {\bibinfo {author} {\bibfnamefont {Y.-Y.}\ \bibnamefont {Chen}}, \bibinfo {author} {\bibfnamefont {K.~Z.}\ \bibnamefont {Hatsagortsyan}}, \bibinfo {author} {\bibfnamefont {C.~H.}\ \bibnamefont {Keitel}},\ and\ \bibinfo {author} {\bibfnamefont {R.}~\bibnamefont {Shaisultanov}},\ }\bibfield  {title} {\bibinfo {title} {Electron spin-and photon polarization-resolved probabilities of strong-field qed processes},\ }\href@noop {} {\bibfield  {journal} {\bibinfo  {journal} {Physical Review D}\ }\textbf {\bibinfo {volume} {105}},\ \bibinfo {pages} {116013} (\bibinfo {year} {2022})}\BibitemShut {NoStop}%
\bibitem [{\citenamefont {Bargmann}\ \emph {et~al.}(1959)\citenamefont {Bargmann}, \citenamefont {Michel},\ and\ \citenamefont {Telegdi}}]{bargmann1959precession}%
  \BibitemOpen
  \bibfield  {author} {\bibinfo {author} {\bibfnamefont {V.}~\bibnamefont {Bargmann}}, \bibinfo {author} {\bibfnamefont {L.}~\bibnamefont {Michel}},\ and\ \bibinfo {author} {\bibfnamefont {V.}~\bibnamefont {Telegdi}},\ }\bibfield  {title} {\bibinfo {title} {Precession of the polarization of particles moving in a homogeneous electromagnetic field},\ }\href@noop {} {\bibfield  {journal} {\bibinfo  {journal} {Physical Review Letters}\ }\textbf {\bibinfo {volume} {2}},\ \bibinfo {pages} {435} (\bibinfo {year} {1959})}\BibitemShut {NoStop}%
\bibitem [{cai()}]{cain_manual}%
  \BibitemOpen
  \href@noop {} {}\bibinfo {howpublished} {\url{https://agenda.linearcollider.org/event/9396/contributions/48972/attachments/37217/58279/CAIN244bmanual.pdf}}\BibitemShut {NoStop}%
\bibitem [{\citenamefont {Abramowicz}\ \emph {et~al.}(2021)\citenamefont {Abramowicz}, \citenamefont {Acosta}, \citenamefont {Altarelli}, \citenamefont {Assmann}, \citenamefont {Bai}, \citenamefont {Behnke}, \citenamefont {Benhammou}, \citenamefont {Blackburn}, \citenamefont {Boogert}, \citenamefont {Borysov} \emph {et~al.}}]{XFEL_LUXE}%
  \BibitemOpen
  \bibfield  {author} {\bibinfo {author} {\bibfnamefont {H.}~\bibnamefont {Abramowicz}}, \bibinfo {author} {\bibfnamefont {U.}~\bibnamefont {Acosta}}, \bibinfo {author} {\bibfnamefont {M.}~\bibnamefont {Altarelli}}, \bibinfo {author} {\bibfnamefont {R.}~\bibnamefont {Assmann}}, \bibinfo {author} {\bibfnamefont {Z.}~\bibnamefont {Bai}}, \bibinfo {author} {\bibfnamefont {T.}~\bibnamefont {Behnke}}, \bibinfo {author} {\bibfnamefont {Y.}~\bibnamefont {Benhammou}}, \bibinfo {author} {\bibfnamefont {T.}~\bibnamefont {Blackburn}}, \bibinfo {author} {\bibfnamefont {S.}~\bibnamefont {Boogert}}, \bibinfo {author} {\bibfnamefont {O.}~\bibnamefont {Borysov}}, \emph {et~al.},\ }\bibfield  {title} {\bibinfo {title} {Conceptual design report for the luxe experiment},\ }\href@noop {} {\bibfield  {journal} {\bibinfo  {journal} {The European Physical Journal Special Topics}\ ,\ \bibinfo {pages} {1}} (\bibinfo {year} {2021})}\BibitemShut {NoStop}%
\bibitem [{SLA()}]{SLAC}%
  \BibitemOpen
  \href@noop {} {}\bibinfo {howpublished} {https://sciencesprings.wordpress.com/2025/03/06/from-the-does-slac-national-accelerator-laboratory-slac-scientists-created-the-most-powerful-ultrashort-electron-beam-in-the-world/}\BibitemShut {NoStop}%
\bibitem [{CER()}]{CERN_LEP}%
  \BibitemOpen
  \href@noop {} {}\bibinfo {howpublished} {\url{https://home.cern/science/accelerators/large-electron-positron-collider}}\BibitemShut {NoStop}%
\bibitem [{\citenamefont {Baier}\ \emph {et~al.}(1973)\citenamefont {Baier}, \citenamefont {Katkov},\ and\ \citenamefont {Fadin}}]{baier1973radiation}%
  \BibitemOpen
  \bibfield  {author} {\bibinfo {author} {\bibfnamefont {V.~N.}\ \bibnamefont {Baier}}, \bibinfo {author} {\bibfnamefont {V.}~\bibnamefont {Katkov}},\ and\ \bibinfo {author} {\bibfnamefont {V.}~\bibnamefont {Fadin}},\ }\bibfield  {title} {\bibinfo {title} {Radiation of relativistic electrons; izluchenie relyativistskikh elektronov},\ }\href@noop {} {\bibfield  {journal} {\bibinfo  {journal} {Unknown}\ } (\bibinfo {year} {1973})}\BibitemShut {NoStop}%
\bibitem [{\citenamefont {Li}\ \emph {et~al.}(2022)\citenamefont {Li}, \citenamefont {Chen}, \citenamefont {Hatsagortsyan},\ and\ \citenamefont {Keitel}}]{li2022helicity}%
  \BibitemOpen
  \bibfield  {author} {\bibinfo {author} {\bibfnamefont {Y.-F.}\ \bibnamefont {Li}}, \bibinfo {author} {\bibfnamefont {Y.-Y.}\ \bibnamefont {Chen}}, \bibinfo {author} {\bibfnamefont {K.~Z.}\ \bibnamefont {Hatsagortsyan}},\ and\ \bibinfo {author} {\bibfnamefont {C.~H.}\ \bibnamefont {Keitel}},\ }\bibfield  {title} {\bibinfo {title} {Helicity transfer in strong laser fields via the electron anomalous magnetic moment},\ }\href@noop {} {\bibfield  {journal} {\bibinfo  {journal} {Physical Review Letters}\ }\textbf {\bibinfo {volume} {128}},\ \bibinfo {pages} {174801} (\bibinfo {year} {2022})}\BibitemShut {NoStop}%
\bibitem [{\citenamefont {Fofanov}\ and\ \citenamefont {et~al.}(2023)}]{Fofanov2023CascadePolarization}%
  \BibitemOpen
  \bibfield  {author} {\bibinfo {author} {\bibfnamefont {A.}~\bibnamefont {Fofanov}}\ and\ \bibinfo {author} {\bibnamefont {et~al.}},\ }\bibfield  {title} {\bibinfo {title} {Cascade of polarized compton scattering and breit–wheeler pair production},\ }\href {https://doi.org/10.1103/PhysRevD.108.116012} {\bibfield  {journal} {\bibinfo  {journal} {Physical Review D}\ }\textbf {\bibinfo {volume} {108}},\ \bibinfo {pages} {116012} (\bibinfo {year} {2023})}\BibitemShut {NoStop}%
\bibitem [{\citenamefont {Aaltonen}\ \emph {et~al.}(2011)\citenamefont {Aaltonen}, \citenamefont {\'Alvarez~Gonz\'alez}, \citenamefont {Amerio},\ and\ \citenamefont {et~al.}}]{Aaltonen2011}%
  \BibitemOpen
  \bibfield  {author} {\bibinfo {author} {\bibfnamefont {T.}~\bibnamefont {Aaltonen}}, \bibinfo {author} {\bibfnamefont {B.}~\bibnamefont {\'Alvarez~Gonz\'alez}}, \bibinfo {author} {\bibfnamefont {S.}~\bibnamefont {Amerio}},\ and\ \bibinfo {author} {\bibnamefont {et~al.}} (\bibinfo {collaboration} {CDF Collaboration}),\ }\bibfield  {title} {\bibinfo {title} {Measurement of polarization and search for $cp$ violation in ${B}_{s}^{0}\ensuremath{\rightarrow}\ensuremath{\phi}\ensuremath{\phi}$ decays},\ }\href {https://doi.org/10.1103/PhysRevLett.107.261802} {\bibfield  {journal} {\bibinfo  {journal} {Phys. Rev. Lett.}\ }\textbf {\bibinfo {volume} {107}},\ \bibinfo {pages} {261802} (\bibinfo {year} {2011})}\BibitemShut {NoStop}%
\bibitem [{\citenamefont {Alexander}\ \emph {et~al.}(2008)\citenamefont {Alexander}, \citenamefont {Barley}, \citenamefont {Batygin}, \citenamefont {Berridge}, \citenamefont {Bharadwaj}, \citenamefont {Bower}, \citenamefont {Bugg}, \citenamefont {Decker}, \citenamefont {Dollan}, \citenamefont {Efremenko}, \citenamefont {Gharibyan}, \citenamefont {Hast}, \citenamefont {Iverson}, \citenamefont {Kolanoski}, \citenamefont {Kovermann}, \citenamefont {Laihem}, \citenamefont {Lohse}, \citenamefont {McDonald}, \citenamefont {Mikhailichenko}, \citenamefont {Moortgat-Pick}, \citenamefont {Pahl}, \citenamefont {Pitthan}, \citenamefont {P\"oschl}, \citenamefont {Reinherz-Aronis}, \citenamefont {Riemann}, \citenamefont {Sch\"alicke}, \citenamefont {Sch\"uler}, \citenamefont {Schweizer}, \citenamefont {Scott}, \citenamefont {Sheppard}, \citenamefont {Stahl}, \citenamefont {Szalata}, \citenamefont {Walz},\ and\ \citenamefont {Weidemann}}]{Alexander2008}%
  \BibitemOpen
  \bibfield  {author} {\bibinfo {author} {\bibfnamefont {G.}~\bibnamefont {Alexander}}, \bibinfo {author} {\bibfnamefont {J.}~\bibnamefont {Barley}}, \bibinfo {author} {\bibfnamefont {Y.}~\bibnamefont {Batygin}}, \bibinfo {author} {\bibfnamefont {S.}~\bibnamefont {Berridge}}, \bibinfo {author} {\bibfnamefont {V.}~\bibnamefont {Bharadwaj}}, \bibinfo {author} {\bibfnamefont {G.}~\bibnamefont {Bower}}, \bibinfo {author} {\bibfnamefont {W.}~\bibnamefont {Bugg}}, \bibinfo {author} {\bibfnamefont {F.-J.}\ \bibnamefont {Decker}}, \bibinfo {author} {\bibfnamefont {R.}~\bibnamefont {Dollan}}, \bibinfo {author} {\bibfnamefont {Y.}~\bibnamefont {Efremenko}}, \bibinfo {author} {\bibfnamefont {V.}~\bibnamefont {Gharibyan}}, \bibinfo {author} {\bibfnamefont {C.}~\bibnamefont {Hast}}, \bibinfo {author} {\bibfnamefont {R.}~\bibnamefont {Iverson}}, \bibinfo {author} {\bibfnamefont {H.}~\bibnamefont {Kolanoski}}, \bibinfo {author} {\bibfnamefont {J.}~\bibnamefont {Kovermann}}, \bibinfo {author} {\bibfnamefont {K.}~\bibnamefont
  {Laihem}}, \bibinfo {author} {\bibfnamefont {T.}~\bibnamefont {Lohse}}, \bibinfo {author} {\bibfnamefont {K.~T.}\ \bibnamefont {McDonald}}, \bibinfo {author} {\bibfnamefont {A.~A.}\ \bibnamefont {Mikhailichenko}}, \bibinfo {author} {\bibfnamefont {G.~A.}\ \bibnamefont {Moortgat-Pick}}, \bibinfo {author} {\bibfnamefont {P.}~\bibnamefont {Pahl}}, \bibinfo {author} {\bibfnamefont {R.}~\bibnamefont {Pitthan}}, \bibinfo {author} {\bibfnamefont {R.}~\bibnamefont {P\"oschl}}, \bibinfo {author} {\bibfnamefont {E.}~\bibnamefont {Reinherz-Aronis}}, \bibinfo {author} {\bibfnamefont {S.}~\bibnamefont {Riemann}}, \bibinfo {author} {\bibfnamefont {A.}~\bibnamefont {Sch\"alicke}}, \bibinfo {author} {\bibfnamefont {K.~P.}\ \bibnamefont {Sch\"uler}}, \bibinfo {author} {\bibfnamefont {T.}~\bibnamefont {Schweizer}}, \bibinfo {author} {\bibfnamefont {D.}~\bibnamefont {Scott}}, \bibinfo {author} {\bibfnamefont {J.~C.}\ \bibnamefont {Sheppard}}, \bibinfo {author} {\bibfnamefont {A.}~\bibnamefont {Stahl}}, \bibinfo {author}
  {\bibfnamefont {Z.~M.}\ \bibnamefont {Szalata}}, \bibinfo {author} {\bibfnamefont {D.}~\bibnamefont {Walz}},\ and\ \bibinfo {author} {\bibfnamefont {A.~W.}\ \bibnamefont {Weidemann}},\ }\bibfield  {title} {\bibinfo {title} {Observation of polarized positrons from an undulator-based source},\ }\href {https://doi.org/10.1103/PhysRevLett.100.210801} {\bibfield  {journal} {\bibinfo  {journal} {Phys. Rev. Lett.}\ }\textbf {\bibinfo {volume} {100}},\ \bibinfo {pages} {210801} (\bibinfo {year} {2008})}\BibitemShut {NoStop}%
\bibitem [{\citenamefont {Xu}\ \emph {et~al.}(2005)\citenamefont {Xu}, \citenamefont {Ho}, \citenamefont {Kong}, \citenamefont {Chen}, \citenamefont {Wang}, \citenamefont {Wang},\ and\ \citenamefont {Lin}}]{Xu2005}%
  \BibitemOpen
  \bibfield  {author} {\bibinfo {author} {\bibfnamefont {J.~J.}\ \bibnamefont {Xu}}, \bibinfo {author} {\bibfnamefont {Y.~K.}\ \bibnamefont {Ho}}, \bibinfo {author} {\bibfnamefont {Q.}~\bibnamefont {Kong}}, \bibinfo {author} {\bibfnamefont {Z.}~\bibnamefont {Chen}}, \bibinfo {author} {\bibfnamefont {P.~X.}\ \bibnamefont {Wang}}, \bibinfo {author} {\bibfnamefont {W.}~\bibnamefont {Wang}},\ and\ \bibinfo {author} {\bibfnamefont {D.}~\bibnamefont {Lin}},\ }\bibfield  {title} {\bibinfo {title} {Properties of electron acceleration by a circularly polarized laser in vacuum},\ }\href {https://doi.org/10.1063/1.2037864} {\bibfield  {journal} {\bibinfo  {journal} {Journal of Applied Physics}\ }\textbf {\bibinfo {volume} {98}},\ \bibinfo {pages} {056105} (\bibinfo {year} {2005})}\BibitemShut {NoStop}%
\bibitem [{\citenamefont {Wang}\ \emph {et~al.}(2001)\citenamefont {Wang}, \citenamefont {Ho}, \citenamefont {Yuan}, \citenamefont {Kong}, \citenamefont {Cao}, \citenamefont {Sessler}, \citenamefont {Esarey},\ and\ \citenamefont {Nishida}}]{Wang2001}%
  \BibitemOpen
  \bibfield  {author} {\bibinfo {author} {\bibfnamefont {P.~X.}\ \bibnamefont {Wang}}, \bibinfo {author} {\bibfnamefont {Y.~K.}\ \bibnamefont {Ho}}, \bibinfo {author} {\bibfnamefont {X.~Q.}\ \bibnamefont {Yuan}}, \bibinfo {author} {\bibfnamefont {Q.}~\bibnamefont {Kong}}, \bibinfo {author} {\bibfnamefont {N.}~\bibnamefont {Cao}}, \bibinfo {author} {\bibfnamefont {A.~M.}\ \bibnamefont {Sessler}}, \bibinfo {author} {\bibfnamefont {E.}~\bibnamefont {Esarey}},\ and\ \bibinfo {author} {\bibfnamefont {Y.}~\bibnamefont {Nishida}},\ }\bibfield  {title} {\bibinfo {title} {Vacuum electron acceleration by an intense laser},\ }\href {https://doi.org/10.1063/1.1359486} {\bibfield  {journal} {\bibinfo  {journal} {Applied Physics Letters}\ }\textbf {\bibinfo {volume} {78}},\ \bibinfo {pages} {2253} (\bibinfo {year} {2001})}\BibitemShut {NoStop}%
\end{thebibliography}%


%

\end{document}